\documentclass{emulateapj}
\usepackage{graphicx}
\usepackage{longtable}    

\begin{document}

\title{High-resolution VLA Imaging of SDSS Stripe 82 at 1.4 GHz}

\shorttitle{High-resolution VLA Imaging of Stripe 82}

\author{J. A. Hodge\altaffilmark{1}, R. H. Becker\altaffilmark{2}}
\affil{University of California, 1 Shields Ave, Davis, CA 95616}

\author{R. L. White}
\affil{Space Telescope Science Institute, 3700 San Martin Drive, Baltimore, MD 21218}

\author{G. T. Richards}
\affil{Drexel University, 3141 Chestnut St., Philadelphia, PA 19104}

\author{G. R. Zeimann}
\affil{University of California, 1 Shields Ave, Davis, CA 95616}

\altaffiltext{1}{Current address: Max Planck Institute for Astronomy, K\"onigstuhl 17, 69117 Heidelberg, Germany.  Email: hodge@mpia.de}
\altaffiltext{2}{Lawrence Livermore National Laboratory, L-413, Livermore, CA 94550}

\begin{abstract}
We present a high-resolution radio survey of the Sloan Digital Sky Survey (SDSS) Southern Equatorial Stripe, a.k.a. Stripe 82.  This 1.4 GHz  survey was conducted with the Very Large Array (VLA) primarily in the A-configuration, with supplemental B-configuration data to increase sensitivity to extended structure.  The survey has an angular resolution of 1.8$^{\prime\prime}$ and achieves a median rms noise of 52 $\mu$Jy beam$^{-1}$ over 92 deg$^2$.  This is the deepest 1.4 GHz survey to achieve this large of an area, filling a gap in the phase space between small, deep and large, shallow surveys.  It also serves as a pilot project for a larger high-resolution survey with the Expanded Very Large Array (EVLA).  We discuss the technical design of the survey and details of the observations, and we outline our method for data reduction.  We present a catalog of 17,969 isolated radio components, for an overall source density of $\sim$195 sources deg$^{-2}$.  The astrometric accuracy of the data is excellent, with an internal check utilizing multiply-observed sources yielding an rms scatter of 0.19$^{\prime\prime}$ in both right ascension and declination.  A comparison to the SDSS DR7 Quasar Catalog further confirms that the astrometry is well tied to the optical reference frame, with mean offsets of 0.02$^{\prime\prime}$ $\pm$ 0.01$^{\prime\prime}$ in right ascension, and 0.01$^{\prime\prime}$ $\pm$ 0.02$^{\prime\prime}$ in declination.  A check of our photometry reveals a small, negative CLEAN-like bias on the level of 35 $\mu$Jy.  We report on the catalog completeness, finding that 97\% of FIRST-detected quasars are recovered in the new Stripe 82 radio catalog, 
while faint, extended sources are more likely to be resolved out by the resolution bias. 
We conclude with a discussion of the optical counterparts to the catalog sources, including 76 newly-detected radio quasars.  The full catalog as well as a search page and cutout server are available online at http://third.ucllnl.org/cgi-bin/stripe82cutout.   






\end{abstract}

\defcitealias{Becker:1995p1065}{BWH95}

\section{INTRODUCTION}
\label{Intro}


The history of radio surveys has played out as the usual tradeoff between depth and width, with most surveys favoring one extreme at the loss of the other.  On one end, large area radio surveys like Faint Images of the Radio Sky at Twenty centimeters \citep[FIRST;][]{Becker:1995p1065} and the NRAO VLA Sky Survey \citep[NVSS;][]{Condon:1998p1324} have been of unquestionable importance to the astronomy community.  These surveys revolutionized the field of radio astronomy with their detailed censuses of the mJy radio population.  Still, they are relatively shallow surveys, with typical rms values of 0.15 mJy beam$^{-1}$ and 0.45 mJy beam$^{-1}$, respectively.  

At the other end of the spectrum are relatively small but deep radio surveys, with pencil beam surveys being the most extreme example.  \citet{Richards:2000p2058} surveyed a 1 deg$^2$ region containing the Hubble Deep Field North (HDF-N) at 1.4 GHz and achieved a completeness limit of 40 $\mu$Jy.  Other examples of deep 1.4 GHz surveys include those done by \citet{Richards:2000p2058, Hopkins:2003p1905, Condon:2003p1901, Bondi:2003p857, Seymour:2004p352, Norris:2005p2030, Norris:2006p2409, Simpson:2006p372, Ivison:2007p1968, Schinnerer:2007p2086, Owen:2008p136, Kellerman:2008p71, Miller:2008p114}; and \citet{Morrison:2010p178}.  The ``widest", arguably, of the deep surveys, the Phoenix Deep Survey \citep[PDS;][]{Hopkins:2003p1905}, covers 4.5 deg$^2$ and has an rms of 12 $\mu$Jy in its most sensitive region.  Such deep surveys have many important applications, including (but not limited to) investigating the dust-unbiased star formation in galaxies, the evolution of the radio-FIR correlation, and the nature of the somewhat controversial sub-mJy population;  however, they can suffer from small number statistics depending on which sources are of interest.  

The future of radio astronomy is bright, or rather, faint, as it turns out.  ASKAP, the Australian Square Kilometer Array Pathfinder, is slated to cover the entire Southern sky with 10$^{\prime\prime}$ resolution to 10 $\mu$Jy rms.  APERTIF, the Westerbork equivalent of ASKAP, will cover the Northern sky with 10$^{\prime\prime}$ resolution to 20 $\mu$Jy rms.  The Expanded Very Large Array (EVLA), which as already begun operations with reduced capabilities, will be 5-20 times more sensitive than the VLA and give point-source sensitivity better than 1 $\mu$Jy between 2-40 GHz.  The Karoo Array Telescope (MeerKat), which South Africa is building as a precursor instrument for the SKA, will provide a southern hemisphere complement to the EVLA from the L-band to the X-band.  

To get around the limitations of the currently available instruments, some have utilized the powerful technique of median stacking.  By taking a population of objects selected at a different wavelength and combining cutouts extracted from a large but shallow survey, it is possible to probe far below the original survey threshold and determine median trends in the data. For example, this technique has been applied to the FIRST survey to study various source populations with median flux densities in the 10s $-$ 100s of $\mu$Jy. \citet{White:2007p1653} studied the radio properties of quasars selected from the Sloan Digital Sky Survey \citep[SDSS;][]{Abazajian:2009p543}.  \citet{deVries:2007p1349} applied the method to SDSS-selected active galactic nuclei (AGN) and star-forming galaxies.  \citet{Hodge:2008p1405} studied the radio emission from galaxies lacking strong optical emission lines, and \citet{Hodge:2009p1413} used median stacking to study low-luminosity AGN in luminous red galaxies (LRGs) out to redshifts of 0.7.  

This method is therefore a way of milking the most out of today's radio surveys, but, like everything, it too has its selection biases.  In this case, the obvious bias is the limitation to samples selected at wavelengths other than the radio.  To study certain faint radio populations without this bias, what is needed is a survey that is both fairly wide and fairly deep.  Here we present a 1.4 GHz VLA survey of the SDSS Southern Equatorial Stripe, also known as ``Stripe 82"\footnote{www.physics.drexel.edu/$\sim$gtr/vla/stripe82/}.  Our survey is intermediate in both size and depth, covering over 90 deg$^2$ to almost three times the depth of FIRST, or an rms of 52 $\mu$Jy.  This makes it the deepest 1.4 GHz survey to cover this much area, and conversely, the widest survey to reach this depth.  

On top of its unique position in depth versus width parameter space, this survey brings with it a significant amount of complementary data.  SDSS Stripe 82 has existing multi-wavelength coverage from the radio to the ultraviolet, the most noteworthy being SDSS imaging two magnitudes deeper than the typical SDSS observations.  Other coverage includes YJHK imaging by the UKIRT Infrared Deep Sky Survey (UKIDSS) as part of its Large Area Survey (LAS), and millimeter imaging by the Atacama Cosmology Telescope (ACT).  The stripe also has multi-\textit{epoch} coverage, with over 100 epochs of SDSS imaging and two previous 1.4 GHz epochs from the FIRST survey.  Finally, our new VLA survey was done with 1.8$^{\prime\prime}$ angular resolution, giving high-resolution radio imaging over the entire 90 deg$^2$ area, as well as serving as a test project for a future FIRST-sized (or larger) high-resolution survey.  

All of these characteristics combine to make this survey a powerful dataset for investigating the extragalactic radio sky; in particular, the AGN.  Radio AGN have typically been divided into radio-loud and radio-quiet objects, with the dividing line generally set near 10$^{25}$ W Hz$^{-1}$, as in \citet{Fanaroff:1974p31}.  The cause for the dichotomy, and even its existence in the first place, has long been subject to debate \citep{Kellermann:1989p1195, Ivezic:2002p2364}.  Recent studies seem to imply that the distribution is formally bimodal, though with a larger radio-intermediate population than previously thought \citep{Cirasuolo:2003p447, Laor:2003arXiv, White:2007p1653}.  

There is also the question of the behavior of the dichotomy as a function of redshift and luminosity.  This is a tricky question to answer, as degeneracy between redshift and luminosity inherent in flux limited surveys may disguise a dependency.  According to \citet{Ivezic:2002p2364}, the fraction of objects classified as radio-loud or radio-quiet does not evolve with either redshift or luminosity.  A more recent study by \citet{Jiang:2007p1999} found that the radio-loud fraction of quasars is a strong function of both redshift and luminosity, decreasing with increasing redshift and decreasing luminosity.  

Related to the evolution of the dichotomy is the evolution of the populations themselves.  It has been known for some time that high-luminosity radio sources show strong cosmic evolution between z $\sim$ 0 and z $\sim$ 2 \citep{Longair:1966p421, Dunlop:1990p19, Willott:2001p536}.  The evolution is more uncertain beyond z $\sim$ 2 \citep{Jarvis:2000p121}, and some studies suggest that it declines \citep[e.g.,][]{Wall:2005p133}.  The evolution of low-luminosity sources is also not well constrained.  Their evolution seems to be weaker than that of powerful radio sources, and some studies imply very weak to almost no evolution at all \citep{Clewley:2004p909, Sadler:2007p211, Donoso:2009p617, Smolcic:2009p24, McAlpine:2010arXiv}.  

Often coming hand-in-hand with measuring radio source evolution is the determination of the radio quasar luminosity function (QLF). Since it is now thought that AGN feedback may be crucial in the formation of massive galaxies \citep[e.g.,][]{Hopkins:2006p1916}, the QLF is one of the key diagnostics in galaxy formation models.  While the QLF has been determined self-consistently from the optical to the hard X-ray \citep{Hopkins:2007p1955}, the radio QLF has only been determined for relatively bright ($>$ 250 mJy) sources \citep{Wall:2005p133}, limiting the ability to constrain galaxy evolution models.

\begin{figure*}
\centering
\includegraphics[scale=0.65]{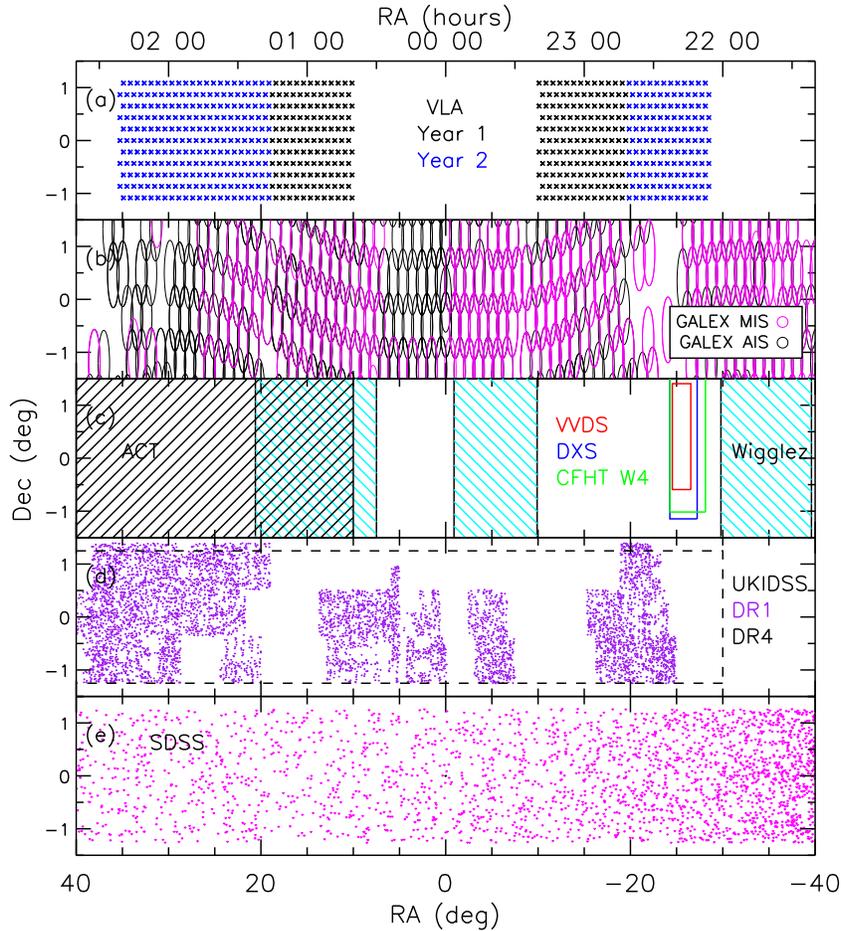}
\caption{Multiwavelength coverage of SDSS Stripe 82.  Panel (a) shows the high resolution VLA coverage of this work, where the `X's mark the individual pointing centers.  Panel (b) shows the GALEX coverage.  Panel (c) shows the initial coverage by ACT.  ACT imaging covering the entire stripe should be public by the end of 2011. Also shown are the areas targeted by the WiggleZ Dark Energy Survey, and the coverage of the VIRMOS-VLT Deep Survey (VVDS), the UKIDSS Deep Extragalactic Survey (DXS), and the CFHT Legacy Survey Wide field 4.  Panel (d) shows the DR1 UKIDSS coverage, with the approximate DR4 coverage indicated by the dashed line.  A random sampling of SDSS photometry is shown in panel (e), where the density increases toward the Galactic Plane.   }
\label{fig:multiwav}
\end{figure*}

The large sample of faint radio sources this survey will provide will help shed light on the issues discussed above and more.  In this paper - the first on the new VLA coverage of the stripe - we describe the basics of the survey and the results of various tests of quality control.  The format of the sections is as follows:  We begin in Section~\ref{design} with a discussion of the technical design of the survey.  In Section~\ref{obs}, we give an account of the observations, followed by an outline of our data reduction steps (Section~\ref{reduction}).  We then present the source catalog (Section~\ref{catalog}), including a brief summary of the source extraction algorithm and definitions of each column in the catalog.  We include a thorough discussion of image characteristics (Section~\ref{characteristics}), including achieved depth, astrometry, photometry, and morphology.  Completeness is discussed in Section~\ref{completeness}.  
We conclude with a section discussing matching to the SDSS and the resulting optical identifications (Section~\ref{optical IDs}).  Where applicable, we assume a flat Friedmann-Lemaitre-Robertson-Walker (commonly FRW) cosmology of a spatially isotropic, homogeneous universe with H$_0$ = 70 km s$^{-1}$ Mpc$^{-1}$, $\Omega_{\Lambda}$ = 0.7, and $\Omega_{M}$ = 0.3.





\section{TECHNICAL DESIGN}
\label{design}

Stripe 82 is the SDSS term for the region of sky that runs along the Celestial Equator in the southern Galactic cap. It was imaged repeatedly by SDSS during the months when the northern Galactic cap was not observable.   As a result, Stripe 82 has much deeper imaging than the rest of SDSS, reaching to g $\sim$24.5 in this roughly 300 square degree area.  

The field also has extensive multi-wavelength coverage.   In the UV, data from the GALEX All-Sky (AIS) and Medium Imaging Surveys (MIS) \citep{Martin:2005p1} exist over much of the field, 
and the UKIRT Infrared Deep Sky Survey \citep[UKIDSS;][]{Lawrence:2007p1599} has targeted Stripe 82 as part of their Large Area Survey (LAS).  
The VISTA Hemisphere Survey \citep{Arnaboldi:2007p28}, reaching J$_{AB}$ $=$ 21.2, has started taking data and will eventually include Stripe 82.  
Millimeter imaging with the Atacama Cosmology Telescope \citep[ACT;][]{Kosowsky:2003p939} reaches a depth of 1 mJy and should be public by the end of 2011.  

Parts of Stripe 82 were also previously covered in the radio by the FIRST survey \citep[hereafter BWH95]{Becker:1995p1065}, which has a typical rms noise of 0.15 mJy beam$^{-1}$ and a resolution of 5".  The large amount of multi-wavelength data made it an obvious choice for deeper radio follow-up.  Our technical goal was therefore to produce a radio survey of Stripe 82 to 3-5 times the depth of the FIRST survey.  The data were taken primarily with the VLA's A-configuration to achieve the highest angular resolution possible.      

As the primary beam is the same size as that used for the FIRST survey, we utilized the same pointing grid as \citetalias{Becker:1995p1065}.  This grid keeps the right ascension spacing fixed within strips of constant declination, and adjusts the declination spacing as necessary so that the sensitivity will be uniform once adjacent fields are coadded.  In this region ($\delta$ $=$ 0$^{\circ}$), that means pointings are separated by 3 min in RA and 13$^{\prime}$ in declination.  

Figure \ref{fig:multiwav} shows the area covered with our high-resolution VLA observations, as well as some of the multi-wavelength coverage.  The `X's in the VLA map in Panel (a) show the individual pointing centers of this work.  We avoided the region around RA $=$ 0 as infrared observations at wavelengths longer than $\sim$5$\mu$m would be affected by the zodiacal light band and contamination by asteroids, which makes it a less useful multi-wavelength region.  Panel (b) shows the GALEX coverage \citep{Martin:2005p1}.  Panel (c) shows the coverage by ACT \citep{Kosowsky:2003p939}, as well as the areas targeted by the WiggleZ Dark Energy Survey \citep{Drinkwater:2010p1429}.  Also shown by the small boxes is coverage by the VIRMOS-VLT Deep Survey \citep[VVDS;][]{LeFevre:2003p18}, the UKIDSS Deep Extragalactic Survey \citep[DXS;][]{Lawrence:2007p1599}, and the CFHT Legacy Survey Wide fields\footnote{http://www.cfht.hawaii.edu/Science/CFHLS/}.  A random sampling of SDSS photometry is shown in panel (d), where the density increases toward the Galactic Plane.  



\section{OBSERVATIONS}
\label{obs}

The data were collected over two VLA cycles (2007-2008, 2008-2009).  The majority of the observations were taken in the A-configuration, but we also obtained B-configuration coverage of the area to improve the sampling of the Fourier (U-V) plane and increase sensitivity to extended structure.  In the first cycle, we were awarded scheduled A-configuration time (Program ID AR646), and the B-configuration time (AR659) was all collected dynamically through the queue.  In the second cycle, we were awarded both the A and B-configuration time as dynamic time (AR685).  We covered Area 1 (Figure \ref{fig:multiwav}) in the A \& B-configurations in 2007-2008, and Area 2 (in A \& B-configurations) in 2008-2009.  Area 1 is made up of 275 pointings, and Area 2 has 374, coming to 649 fields, and 92 deg$^{2}$ covered in total.  


Due to various issues with the data collection and the use of dynamic time for all but the 2007-2008 A-configuration, the total time on each field varies considerably.  For the 2007-2008 scheduled A-configuration time (AR646), each field has anywhere from five to ten 284-second observations (time on source, or ``dwell" time) devoted to it, or 1420--2840 seconds.  For the 2007-2008 B-configuration time (dynamic, AR659), we obtained at least one 284-second observation of each field, and as many as four 284-second observations (1136 seconds total dwell) near $\delta$ $=$ 0$^{\circ}$ in Area 1.  

For the 2008-2009 cycle, we were awarded dynamic time in both the A and B-configurations.  For the A-configuration, we divided our allotted 188 hours such that every field was observed for at least three 495-second intervals, for a total of 1485 seconds (dwell) per field.  Some fields were affected by WIDAR testing, which utilized the easternmost antenna.  Since good resolution is one of our goals for this project, we have also re-observed those fields during non-WIDAR time.  For the B-configuration, we were not able to use all of our time due to the popularity of our LST range during this cycle.  Some data had to be collected during the BnA-configuration.  In the end, we succeeded in obtaining at least 220 seconds (dwell) on every field. 


To minimize bandwidth smearing, where sources away from the beam center are radially blurred due to the finite receiving bandwidth, our observations utilize bandwidth synthesis.  We observed in spectral line mode with 7 frequency channels of 3.125 MHz each.  Correlator limitations then set the bandwidths of the 2 IFs (for the two circular polarizations) at 25 MHz, and these were centered at 1465 and 1385 MHz.  Our chosen integration time was 3 seconds.  We used a number of phase calibrators spread out over the area in order to minimize slew time, and we typically observed a phase calibrator every 30-60 minutes for 40-90 seconds.  We tied the observations to a flux scale using 3C48 as our flux calibrator, which was observed once during every observation block.

\section{DATA REDUCTION}
\label{reduction}

We reduced the data using the Astronomical Imaging Processing System (AIPS) software.  As the data were taken during the transition from the VLA to the EVLA, we created a new Channel 0 dataset from the line data in order to avoid closure errors.  We first applied baseline corrections and ran UVFIX to correct post-MODCOMP uvw errors: some of our early data had u'w, v's, and w's written with the wrong sign, and there was also a problem with the online calculation of negative u's, v's, and w's.  Then we flagged the data by hand with CLIPM and UVFLG to remove any problems due to interference, cross-talk between antennas, correlator failures, or bad antennas.  All the data went through four rounds of self-calibration (two phase-only, and two phase and amplitude).


We imaged the data with IMAGR using multiple facets and three-dimensional techniques.  We used 265 facets with 1024 0.3$^{\prime\prime}$ pixels per side. The fields used for cleaning were thus very large, and we did not find it necessary to take into account sources outside the field.  The IFs were imaged together in all fields except a small number of problem fields where the resultant rms was significantly higher than the survey median.  To keep as uniform as possible, we cleaned each field down to 0.2 mJy. There is also a trade-off between sensitivity and resolution.  Due to the declination of this stripe, the VLA beam is significantly extended in the N-S direction from the nominal A-configuration resolution of 1.4$^{\prime\prime}$. To maintain a circular beam while using natural weighting, we found that the resolution degraded to 2.5$^{\prime\prime}$.  In the final round of imaging, we therefore use uniform weighting with robust$=$0 and a circular beam of 1.8$^{\prime\prime}$.  The facets were stitched back together with FLATN using a cutoff radius of 22.3$^{\prime}$, and the final images are 4650 $\times$ 4650 0.6$^{\prime\prime}$ pixels.  These images, which were made from individual pointings, are the so-called ``grid" images.  The typical rms for the grid images is 52 $\mu$Jy (see Section \ref{depth}).  


To produce the final maps, we co-added the individual grid images using an adaptation of the pipeline produced for the FIRST survey by BWH95.  This step increases the S/N and improves sensitivity away from the beam center.  The pipeline searches for fields in the vicinity of the map center, truncates each where the primary beam sensitivity falls below 18\%, weights by the square of the primary beam response, and sums the fields.  Due to overlapping coverage, we produce co-added maps for only every other declination strip.  The resultant maps are 4650 $\times$ 4650 pixels, or 46.5$^{\prime}$ $\times$ 46.5$^{\prime}$ in extent, with 0.6$^{\prime\prime}$ pixels.

\begin{table*}[ht]
  \caption{Excerpt from the VLA Stripe 82 Survey Catalog} 
\scalebox{0.8}
 {
\begin{tabular}{llrrrrrrrrrrrrrcr}	
  \hline\hline \\[-2ex]
  & & & & & & & & & & & & & \multicolumn{4}{c}{\textbf{----SDSS Matches----}} \\
 \multicolumn{1}{c}{\textbf{RA}} & \multicolumn{1}{c}{\textbf{Dec}} & \multicolumn{1}{c}{\textbf{P(S)}} &  \multicolumn{1}{c}{\textbf{F$_{\rm peak}$}} & \multicolumn{1}{c}{\textbf{F$_{\rm int}$}} & \multicolumn{1}{c}{\textbf{RMS}} & \multicolumn{1}{c}{\textbf{Maj}} & \multicolumn{1}{c}{\textbf{Min}} & \multicolumn{1}{c}{\textbf{PA}} & \multicolumn{1}{c}{\textbf{fMaj}} & \multicolumn{1}{c}{\textbf{fMin}} & \multicolumn{1}{c}{\textbf{fPA}} & \multicolumn{1}{c}{\textbf{Field}} & \boldmath{\#} & \textbf{Sep} & \textbf{i} & \textbf{Cl}  \\
  \multicolumn{1}{c}{\textbf{(J2000)}} &  \multicolumn{1}{c}{\textbf{(J2000)}} &  &  \multicolumn{1}{c}{\textbf{(mJy)}} &  \multicolumn{1}{c}{\textbf{(mJy)}} &  \multicolumn{1}{c}{\textbf{(mJy)}} &  \multicolumn{1}{c}{\textbf{($^{\prime\prime}$)}} &   \multicolumn{1}{c}{\textbf{($^{\prime\prime}$)}} &  \multicolumn{1}{c}{\textbf{(deg)}} &  \multicolumn{1}{c}{\textbf{($^{\prime\prime}$)}} &   \multicolumn{1}{c}{\textbf{($^{\prime\prime}$)}} &  \multicolumn{1}{c}{\textbf{(deg)}} & & & \multicolumn{1}{c}{\textbf{($^{\prime\prime}$)}} & \multicolumn{1}{c}{\textbf{(mag)}} &  \\ [0.5ex] \hline\hline
 \\[-1.8ex]
22 26 25.132 & +00 25 09.26 & 0.317  &   0.52  &    0.29     &  0.092 &  0.00   &  0.00 & 46.3  &  1.60 &  1.15 & 46.3  & 22255+00390H & 6  & 0.27  & 20.69 & g \\
22 20 19.710 & +00 25 08.57 & 0.276  &   0.53  &    0.45     &  0.085 &  0.00   &  0.00 & 180.0 &  1.76 &  1.59 & 180.0 & 22195+00130H & 5  & 0.19  & 19.91 & g \\
02 06 42.812 & +00 25 08.36 & 0.200  &   0.97  &    1.32     &  0.063 &  1.14   &  1.01 &  0.0  &  2.13 &  2.07 &  0.0  & 02075+00130H & 3  & 0.08  & 16.58 & g \\
22 11 36.231 & +00 25 07.48 & 0.528  &   0.33  &    0.33     &  0.065 &  0.45   &  0.00 &  0.0  &  1.86 &  1.75 &  0.0  & 22105+00130H & 6  & 4.33  & 20.88 & s \\
22 51 53.502 & +00 25 06.52 & 0.822  &   0.50  &    4.84     &  0.096 &  5.78   &  4.91 & 80.2  &  6.05 &  5.23 & 80.2  & 22525+00130H & 0  & 99.00 & 99.00 & - \\
01 28 00.209 & +00 25 06.30 & 0.862  &   0.52  &    0.63     &  0.093 &  2.00   &  0.00 &  6.1  &  2.69 &  1.46 &  6.1  & 01285+00130H & 2  & 5.55  & 21.64 & s \\
23 11 14.065 & +00 25 05.97 & 0.751  &   0.32  &    0.51     &  0.057 &  1.50   &  1.28 & 180.0 &  2.34 &  2.21 & 180.0 & 23105+00390H & 1  & 6.92  & 20.91 & g \\
22 20 47.692 & +00 25 05.53 & 0.049  &   1.19  &    1.84     &  0.107 &  1.34   &  1.32 & 178.2 &  2.24 &  2.23 & 178.2 & 22195+00130H & 1  & 7.00  & 22.62 & g \\
22 36 12.430 & +00 25 05.44 & 0.415  &   0.85  &    1.12     &  0.090 &  1.38   &  0.54 & 0.8   &  2.27 &  1.88 &  0.8  & 22375+00130H & 25 & 3.50  & 20.80 & g \\
22 27 21.638 & +00 25 04.74 & 0.205  &   0.39  &    1.32     &  0.065 &  2.98   &  2.55 & 112.0 &  3.48 &  3.12 & 112.0 & 22285+00130H & 0  & 99.00 & 99.00 & - \\
23 07 31.374 & +00 25 04.14 & 0.284  &   0.38  &    0.65     &  0.065 &  1.55   &  1.50 & 180.0 &  2.37 &  2.34 & 180.0 & 23075+00390H & 2  & 1.62  & 17.58 & g \\
01 57 23.646 & +00 25 03.98 & 0.252  &   0.63  &    1.15     &  0.091 &  1.67   &  1.56 & 175.0 &  2.45 &  2.39 & 175.0 & 01585+00130H & 7  & 0.11  & 15.64 & g \\
00 50 16.367 & +00 25 03.69 & 0.317  &   0.33  &    0.52     &  0.065 &  1.36   &  1.35 &  0.0  &  2.25 &  2.25 & 180.0 & 00495+00130H & 2  & 0.26  & 20.83 & g \\
01 17 51.785 & +00 25 03.55 & 0.014  &   1.49  &    3.31     &  0.054 &  2.65   &  1.34 & 11.2  &  3.21 &  2.25 & 11.2  & 01165+00130H & 5  & 3.65  & 20.92 & g \\
01 17 51.452 & +00 25 02.73 & 0.014  &   2.63  &    4.00     &  0.054 &  1.67   &  0.88 & 138.9 &  2.45 &  2.01 & 138.9 & 01165+00130H & 4  & 1.89  & 20.02 & g \\
22 26 17.340 & +00 25 01.88 & 0.208  &   0.45  &    0.42     &  0.078 &  0.54   &  0.00 & 180.0 &  1.88 &  1.63 & 180.0 & 22255+00130H & 0  & 99.00 & 99.00 & - \\
22 44 02.193 & +00 25 01.25 & 0.149  &   0.85  &   11.61     &  0.073 &  6.88   &  5.95 &  5.1  &  7.11 &  6.22 &  5.1  & 22435+00130H & 4  & 7.43  & 22.26 & g \\
23 15 43.210 & +00 25 01.03 & 0.797  &   0.31  &    2.02     &  0.058 &  4.42   &  4.06 & 178.2 &  4.78 &  4.44 & 178.2 & 23165+00130H & 19 & 0.16  & 13.65 & g \\
22 46 37.766 & +00 25 00.78 & 0.222  &   0.71  &    3.72     &  0.064 &  3.86   &  3.55 & 105.8 &  4.26 &  3.98 & 105.8 & 22465+00130H & 7  & 3.17  & 21.51 & g \\
22 27 21.717 & +00 25 00.77 & 0.014  &   9.64  &   11.48     &  0.065 &  0.80   &  0.77 & 174.8 &  1.97 &  1.96 & 174.8 & 22285+00130H & 2  & 1.02  & 19.70 & g \\
02 21 04.213 & +00 24 59.89 & 0.014  &   4.88  &    5.22     &  0.060 &  0.56   &  0.38 &  0.0  &  1.88 &  1.84 &  0.0  & 02210+00260H & 4  & 0.15  & 21.66 & g \\
00 43 32.712 & +00 24 59.84 & 0.034  &  76.14  &   104.13    &  0.698 &  1.44   &  0.67 & 95.4  &  2.31 &  1.92 & 95.4  & 00435+00130H & 3  & 0.04  & 19.11 & s \\
01 45 31.568 & +00 24 59.50 & 0.188  &   1.78  &    2.00     &  0.104 &  1.01   &  0.00 &  6.9  &  2.06 &  1.77 &  6.9  & 01465+00130H & 2  & 0.09  & 20.12 & s \\
22 27 21.947 & +00 24 59.43 & 0.014  &   3.80  &    7.91     &  0.065 &  2.24   &  1.50 & 164.8 &  2.88 &  2.34 & 164.8 & 22285+00130H & 0  & 99.00 & 99.00 & - \\
22 20 47.714 & +00 24 58.85 & 0.014  &   5.95  &    6.45     &  0.107 &  0.53   &  0.51 & 179.9 &  1.88 &  1.87 & 179.9 & 22195+00130H & 1  & 7.10  & 20.79 & g \\
22 44 02.153 & +00 24 58.79 & 0.368  &   0.60  &   17.63     &  0.073 & 10.82   &  8.50 & 67.8  & 10.97 &  8.69 & 67.8  & 22435+00130H & 0  & 99.00 & 99.00 & - \\
02 21 26.327 & +00 24 57.02 & 0.075  &   2.43  &    3.49     &  0.062 &  1.64   &  0.65 & 171.1 &  2.44 &  1.91 & 171.1 & 02210+00260H & 2  & 2.22  & 19.02 & s \\
\\ [-1.8ex]\hline\hline   \\[-1.9ex]
  \end{tabular}  
   \footnotetext{NOTES. See Section 5.2 for column definitions.}
   }
  \label{tab:catalog}
 \end{table*}

\section{THE CATALOG}
\label{catalog}
\subsection{Source Extraction}

To create the catalog, we used HAPPY, the source-finding algorithm which was developed initially for use by the FIRST survey \citet{White:1997p2118}.  HAPPY searches an image for pixels exceeding a user-defined threshold, and it places rectangles around contiguous regions of such pixels to form ``islands".  Each island is then fit with up to four two-dimensional Guassian components, which must pass several criteria for acceptance.  The end result is a list of elliptical Gaussian components.  More details on HAPPY can be found in \citet{White:1997p2118}.  

HAPPY requires a noise threshold to be specified, which we compute as a function of position in a coverage map.  To create the map, we combine the rms HAPPY computes by looking at a histogram of pixel values for each grid image with a model for the elevated noise around bright sources, with the noise being much higher in the north-south direction.  This additional step is necessary in order to ensure the strong sidelobes present around many of the bright sources in this survey do not end up in the source catalog.  We then add the contributing noise maps at each position, weighting them in the same way we weight the grid images that make up the coadded images.  We set HAPPY's noise threshold as five times the rms found in this coverage map at the position of interest.  To view the coverage map for Stripe 82, see Section~\ref{depth}.  

Since the coadded images have some overlap, a catalog produced in this way will have duplicate sources.  Identifying the duplicates is generally straightforward, since the images that contribute at any sky position are exactly the same in the coadded images that cover that position.  The only difference is that the pixel grid centers will be slightly different.  For the vast majority of sources, the fitted parameters are almost identical, and they agree much more closely than the actual uncertainty in the parameters.  

There may be cases, however, where a source gets decomposed into multiple components, and the source fits are therefore different in different fields.  To identify groups of duplicate sources and select which sources to retain from the matched groups, we utilize the following algorithm:    
We first search for matches within 1.7$^{\prime\prime}$ of each source, where 1.7$^{\prime\prime}$ is chosen because it is approximately the size of the PSF FWHM.  Groups may be chained together by links of $\leq$1.7$^{\prime\prime}$, so the total group size may be larger than this if there are multiple links.  Sources that are separated by more than 1.7$^{\prime\prime}$ are resolved into separate components.  We then find all coadded fields that overlap the mean RA/Dec position for the group, and we identify the field where the group is closest to the center as the best field.  We then keep only the sources that come from that best field, discarding duplicates from fields other than the best one.  This is different than the approach used by FIRST and outlined in \citet{White:1997p2118}, and it is applied to all sources, not just those with duplicates.  The final catalog should therefore only contain sources from fields that are identified as the best field at that source position.  Applying this process to the Stripe 82 data, we produce a final catalog of 17969 radio sources.  An example page from the catalog is shown in Table \ref{tab:catalog}.  


\subsection{Column Definitions}

The columns are well-defined on the FIRST survey website\footnote{http://sundog.stsci.edu}, but we give a brief summary here.

Columns (1) and (2) - Right ascension and declination of the source in J2000.  The positional errors are a function of source brightness, size, and noise in the map.  They are best found using a simple rule-of-thumb approach, as the HAPPY-derived errors tend to be underestimated.  An empirical equation for the accuracy at 90\% confidence is $f$Size$\times (1/SNR + 1/20)$, where $f$Size is the fitted major or minor axis size, and $SNR$ is the signal to noise ratio \citep{White:1997p2118}.  Systematic errors are smaller than 0.05$^{\prime\prime}$.

Column (3) - P(S) indicates the probability that the source is spurious, most likely because it is a sidelobe of a nearby bright source.  Low values of P(S) mean the source is unlikely to be spurious.  The probabilities are computed using an algorithm based on multiple voting oblique decision tree classifiers, which were trained on deep VLA fields.  The algorithm was developed for the FIRST survey and will be described in more detail in a future paper on the final FIRST catalog.  Note that the algorithm is optimized for the FIRST survey, whereas the RMS computation for this catalog has changed significantly.  The values of P(S) are therefore not very reliable for this catalog.

Column (4) - Fpeak is the peak flux density in mJy beam$^{-1}$ derived by fitting an elliptical Gaussian model to the source.  The uncertainty is given by the rms noise at the source position.

Column (5) - Fint is the integrated flux density in mJy derived from the elliptical Gaussian model fit.  The uncertainty in Fint can be considerably greater than that of Fpeak depending on source size and morphology.  An expression to estimate the uncertainty can be found in \citet{Schinnerer:2004p128}.  For point sources, the relative uncertainty ($\sigma_{I}/I$) reduces to:
\begin{eqnarray*}
\frac{\sigma_{I}}{I} = \sqrt{{2.5 \frac{\sigma^2}{I^2}} + {0.01}^{2} }.
\end{eqnarray*}



Column (6) - RMS is the local noise estimate at the source position in mJy.  The RMS is calculated by combining the measured noise from all grid images contributing to the coadded map at the source position.  

Columns (7), (8) and (9) - Maj, Min, and PA give the major and minor axes (FWHM in arcsec) and position angle (degrees east of north) derived from the elliptical Gaussian model for the source.  Maj and Min have had the elliptical Gaussian point-spread function deconvolved.  Noise can cause the fitted values of major and minor axis prior to deconvolution to be smaller than the beam, and the deconvolved size is given as zero in that case.  The uncertainties in the deconvolved sizes depend on both brightness and size.  

Columns (10), (11) and (12) - fMaj, fMin, and fPA give the major and minor axes (FWHM in arcsec) and position angle (degrees east of north) derived from the elliptical Gaussian model for the source.  These are the fitted sizes before deconvolution of the 1.8" circular clean beam.

Column (13) - The field name is the name of the coadded image containing the source.  Note that the field name encodes the center of the field; field hhmmm+ddmmm is centered at RA = hh mm.m, Dec = +dd mm.m.  The letter appended to the field name indicates the last catalog release in which the image was modified.  

Columns (14), (15), (16), and (17) give information on optical counterparts from the SDSS DR6, which were found using the CasJobs web interface.  Column (14) gives the number of matches within a radius of 8$^{\prime\prime}$.  For the closest match, Columns (15) - (17) give (respectively) the separation in arcsec, magnitude in the SDSS i-band, and morphological class, where \textbf{s}$=$stellar and \textbf{g}$=$nonstellar/galaxy.

\section{IMAGE ANALYSIS}
\label{analysis}

This section introduces two effects, bandwidth smearing and resolution bias, which will influence the derived catalog properties discussed in the next section (Section \ref{characteristics}) in complex ways.  For further discussion of these effects and possible correction schemes, the reader is referred to (for e.g.) \citet{Huynh:2005p1960}, \citet{Prandoni:2001p2045}, and \citet{Bondi:2008p1129}.

\begin{figure}
\centering
\includegraphics[scale=0.43]{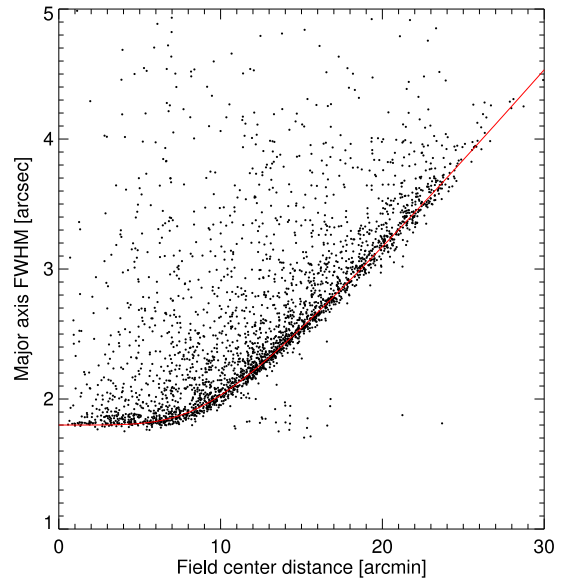}
\caption{Fitted major axis size versus distance from field center for all grid image sources with S$_{peak}$ $>$ 5mJy.}
\label{fig:bwbeam6}
\end{figure}

\subsection{Bandwidth Smearing}
\label{bwsmear}

The effect of bandwidth smearing is to radially-smear the source flux, thus causing objects to appear more extended and the peak flux to be underestimated.  
To calculate the effect of bandwidth smearing on the beam size, we plot measured major axis size versus distance from the field center for all the grid image sources with S$_{peak}$ $>$ 5 mJy beam$^{-1}$ (Figure \ref{fig:bwbeam6}).  We use only bright sources to ensure the results are not affected by spurious sources at fainter flux densities.  Some sources are resolved, which produces a scattering of points toward high values of the major axis, but there is a clear lower bound that is defined by point sources.   
We fit a function to this ridge (red line) and used it to compute the weighted, smeared beam minor axis versus major axis for all the catalog sources.  We do this by calculating the moments (x$^2$, y$^2$, xy) for the point spread function in each grid image including the effect of bandwidth smearing, which stretches the sources in the radial direction.  Depending on where the source falls in the pattern of grid images, there may be several contributing images with stretching oriented in different directions.  We calculate the weighted sum of all these moments (in the same way we weighted the grid images to produce the coadded image), and convert back to a Gaussian model with the major axis, minor axis, and position angle computed from the moments.

\begin{figure}
\centering
\includegraphics[scale=0.43]{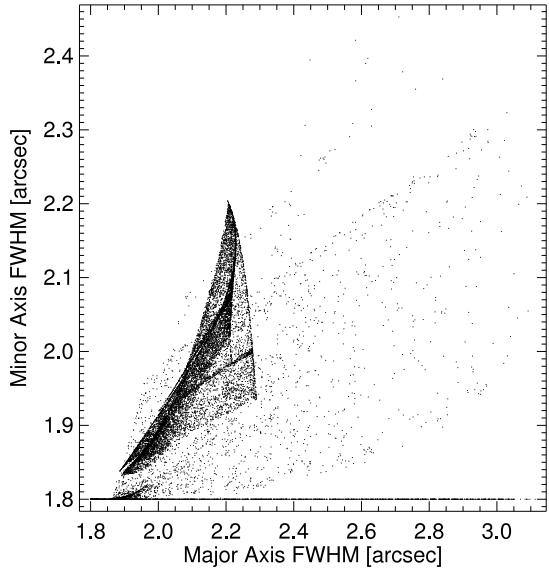}
\caption{The measured weighted, smeared beam minor and major axes for all the catalog sources.  Without bandwidth smearing, all sources would be at the position (1.8$^{\prime\prime}$, 1.8$^{\prime\prime}$).}
\label{fig:bwbeam1}
\end{figure}

\begin{figure}
\centering
\includegraphics[scale=0.43]{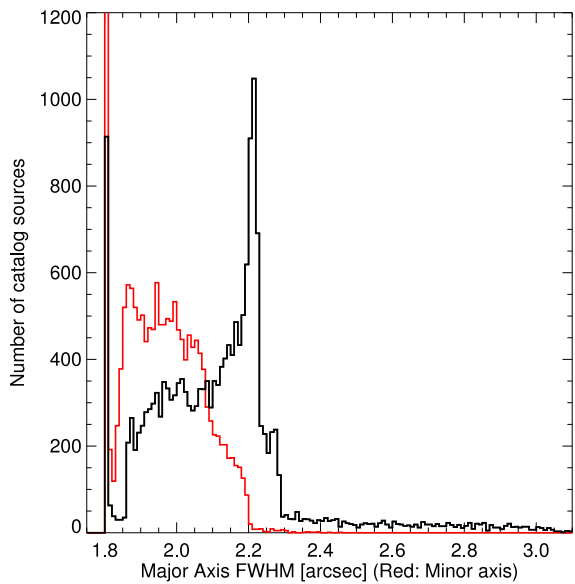}
\caption{The histogram of major and minor axis sizes for the points in Figure \ref{fig:bwbeam1}.  The median major axis is 2.13$^{\prime\prime}$, and the median minor axis is 1.95$^{\prime\prime}$.}
\label{fig:bwbeam3}
\end{figure}

Figure \ref{fig:bwbeam1} shows the weighted, smeared beam minor axis versus major axis for all the catalog sources.  Without the effect of bandwidth smearing, all points would be at the position (1.8, 1.8).  Due to bandwidth smearing, the bulk of the points have slightly large major and minor axes.  (Note that since the beam may be smeared in different directions in the different grid images, the minor axis is also typically larger than 1.8$^{\prime\prime}$).  Complex structure is visible in the Figure, which traces the geometry of the overlapping regions between fields.  There is also an obvious scattering of points which extend out to even larger beam sizes (major axis $\sim$3$^{\prime\prime}$).  These points represent sources at the edge of our coverage, where there are no observations close to field-center, and all observations are therefore far off axis.  

\begin{figure}
\centering
\includegraphics[scale=0.49]{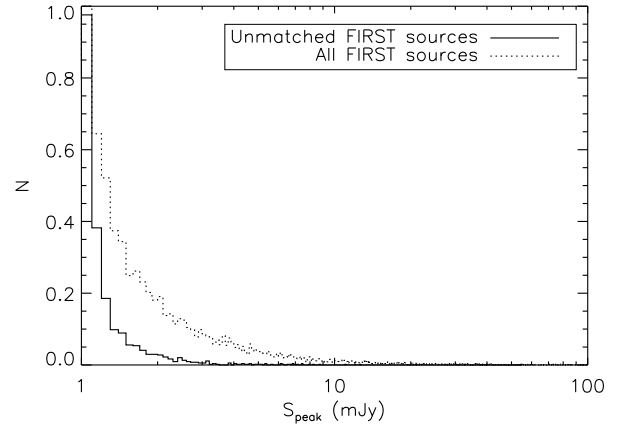}
\caption{The solid histogram shows the distribution of values of peak flux density for FIRST sources that have no match in our new Stripe 82 radio catalog.  The dotted histogram shows the distribution of values of peak flux density for all FIRST sources in the area, including those with matched sources in the new Stripe 82 catalog.  Both histograms have been normalized to a peak value of 1.0.}
\label{fig:unmatchedFIRSTfluxes}
\end{figure}

Figure \ref{fig:bwbeam3} shows the histogram of major and minor axis sizes for the distribution of points in Figure \ref{fig:bwbeam1}.  The median major and minor axis sizes are 2.13$^{\prime\prime}$ and 1.95$^{\prime\prime}$, respectively, and the means are essentially identical to the medians.  There is a well-defined peak at a major axis of 2.2$^{\prime\prime}$.  There is also a peak at exactly 1.8$^{\prime\prime}$ corresponding to sources at the centers of grid images where there are no overlapping grid images, and therefore no bandwidth smearing.  The current catalog version does not include a correction for the effect of bandwidth smearing.  One possible fix would be to degrade everything to the same resolution by smearing the beams with a kernel of the appropriate size prior to coaddition of the grid images.  It is important to keep in mind that bandwidth smearing will affect all of the survey characteristics discussed in Section \ref{characteristics}.

\begin{figure}
\centering
\includegraphics[scale=0.49]{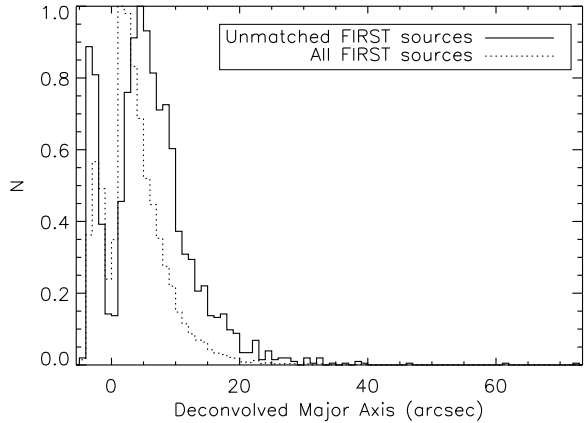}
\caption{The solid histogram shows the distribution of deconvolved major axis sizes for FIRST sources that have no match in our new Stripe 82 catalog.  The dotted histogram shows the distribution of sizes for all FIRST sources in the area, including those with matched sources in the new Stripe 82 catalog.  Both histograms have been normalized to a peak value of 1.0. Negative values of deconvolved major axis indicate sources that have a fitted major axis size smaller than the beam and likely are due to sidelobes, which often appear elongated in one direction and narrow. }
\label{fig:unmatchedFIRSTsizes}
\end{figure}

\begin{figure*}
\centering
\includegraphics[scale=0.7]{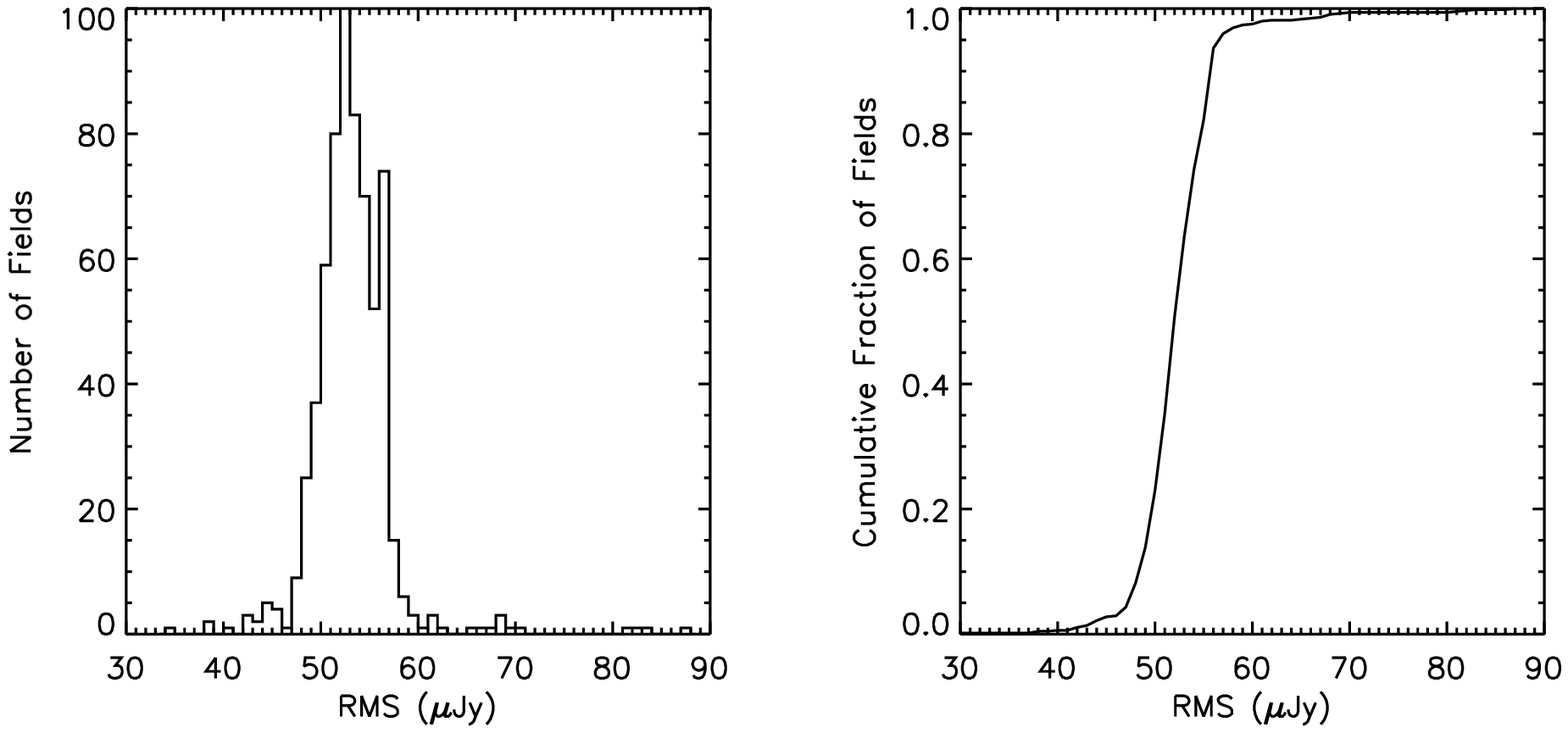}
\caption{Left: a histogram showing the rms noise levels for the 649 grid images that make up our co-added maps.  Right: the cumulative fraction of fields as a function of rms noise. }
\label{fig:rmshistcdf}
\end{figure*}

\subsection{Resolution Bias}
\label{resbias}

Resolution bias is due to the fact that resolved sources with integrated fluxes over the catalog cut will not make it into the catalog if their peak flux is below the cut.  As the Stripe 82 survey was conducted at high resolution, we therefore expect to see a large effect from the resolution bias\footnote{One might ask why the catalog is not instead constructed so that integrated flux is the limiting criterion; however, as interferometers will always ``resolve out" flux from extended structures with angular sizes larger than the largest available resolution element of the array, a catalog constructed in this fashion would suffer from its own incompleteness.}.  One simple way to estimate the effect of the resolution bias is to compare our catalog to the lower-resolution FIRST survey.  By looking at how many FIRST sources are undetected in the Stripe 82 catalog, we can get an estimate of the number of sources that are being ``resolved out" between the FIRST catalog and the Stripe 82 catalog.  

Of the 11008 FIRST sources in the Stripe 82 area, 8635 have one or more match within 5$^{\prime\prime}$ in the Stripe 82 catalog, leaving 2373 FIRST sources without a Stripe 82 counterpart (22\%).  A (normalized) histogram showing the distribution of values of peak flux density for these FIRST sources is shown in Figure \ref{fig:unmatchedFIRSTfluxes} (solid line).  
For comparison, in the same figure we also show the distribution of peak flux density values for all of the FIRST sources (dotted line).  The distribution of unmatched FIRST sources is clearly skewed toward low values of peak flux density.  This result is consistent with what we would expect if resolution effects are at play, since fainter sources are more easily resolved out.

To determine whether the unmatched FIRST sources tend to be extended, we plot the normalized distribution of deconvolved major axis sizes compared to that of all FIRST sources in Figure \ref{fig:unmatchedFIRSTsizes}.  
The unmatched sources appear to be slightly skewed toward larger sizes.  As a precaution, we repeated the matching between the FIRST and new Stripe 82 catalogs, but using a matching radius equal to one-half of the fitted major axis size instead.  The matching radius thus scales with source size, ensuring that we are not missing a valid population of resolved but legitimate sources by not searching far enough from the FIRST source center. This match still resulted in $\sim$21\% 
of the FIRST sources lacking a Stripe 82 counterpart.  We conclude that, as expected, extended sources are more likely to be resolved out.  The resolution bias will affect the photometry, completeness, and morphology of the catalog, which will be discussed in the following section.

\begin{figure*}
\centering
\includegraphics[scale=0.7]{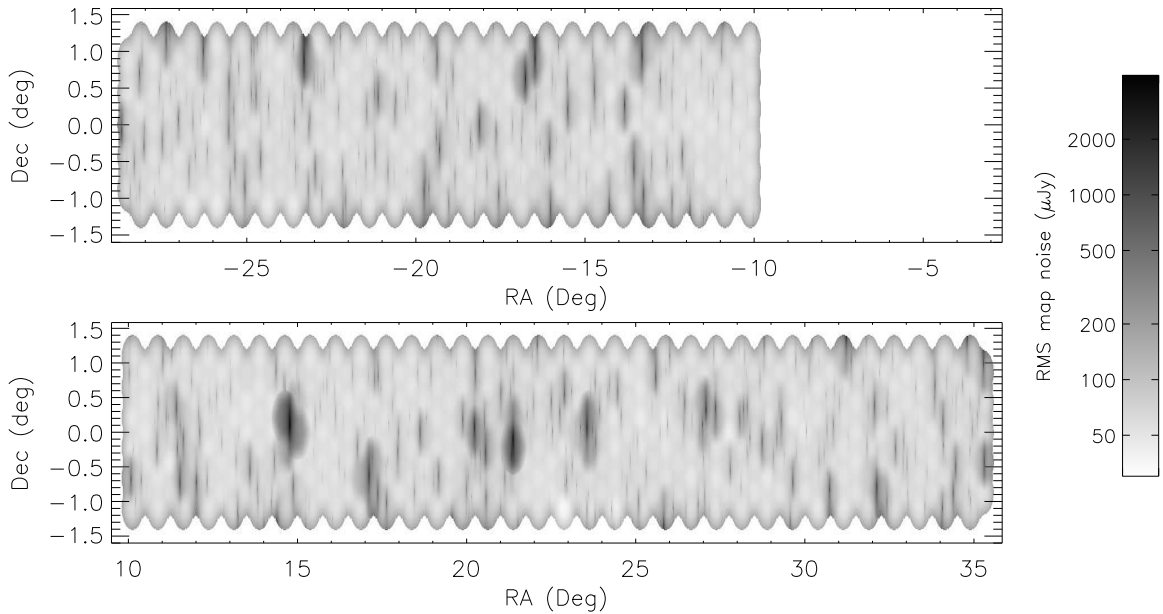}
\caption{Sensitivity map showing the rms noise achieved across the surveyed area.  Note that the x and y axes are not to scale.}
\label{fig:coveragemap}
\end{figure*}

\section{SURVEY CHARACTERISTICS}
\label{characteristics}
\subsection{Depth}
\label{depth}

Figure \ref{fig:rmshistcdf} shows a histogram of the rms noise levels for the 649 grid images that make up our co-added maps, as well as the cumulative fraction of fields as a function of rms noise.  The median rms is 52 $\mu$Jy beam$^{-1}$, and 97\% of the fields have rms noise levels $<$ 60 $\mu$Jy beam$^{-1}$.  
We therefore achieve our target rms noise of three times the depth of FIRST, or $\sim$50 $\mu$Jy beam$^{-1}$.  We note that there are a few images where the noise is determined by the dynamic range, and they are compromised by bright sources.  We also note that in the immediate vicinity of bright sources, the noise is significantly worse than the total image rms.  This will be discussed below.

A coverage map, showing the rms sensitivity achieved in the co-added images, is shown in Figure \ref{fig:coveragemap}.  
The rms sensitivity is shown for the entire region surveyed with a resolution of 1$^{\prime}$ $\times$ 1$^{\prime}$, which is only for presentation purposes, as the catalog is created by computing the rms at the exact position of each source.  
Note that the x and y axes are not to scale, as the surveyed strip is very thin in Declination.  
To make the map, we created a matrix of the desired resolution, and for each position in the matrix, we determine which grid images contributed to the co-added map at that position.  We then took the rms values of the grid images and convolved them with a model for the noise around bright sources.  This was necessary in order to exclude sidelobes from the final source catalog.

Finally, we combined these values with the same weighting used to co-add the grid images.  The resultant rms sensitivity values per 1 arcmin$^2$ region range from 35 $\mu$Jy beam$^{-1}$ to 4.4 mJy beam$^{-1}$.  The high end is due to areas in the immediate vicinity of very bright sources.  The sensitivity map is used to establish the 5$\sigma$ threshold for catalog sources.

\begin{figure}
\centering
\includegraphics[scale=0.6]{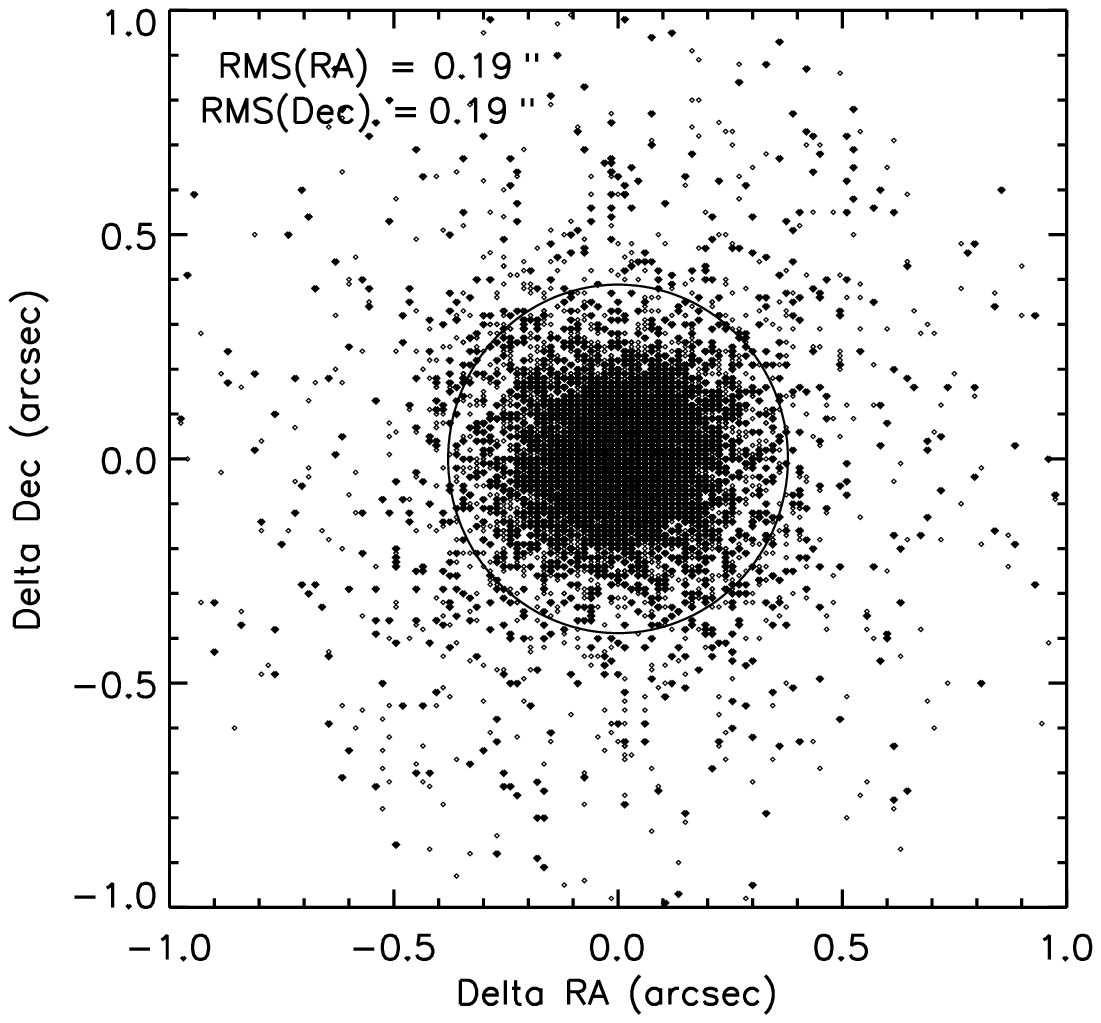}
\caption{The positional offsets for multiply-observed sources with S $>$ 1 mJy.  The ellipse shows two times the rms scatter in right ascension and declination.  Our high-resolution survey achieves an astrometric accuracy approximately 2.5 times that of FIRST, with an rms scatter of 0.19$^{\prime\prime}$ in right ascension and declination. }
\label{fig:posncheck}
\end{figure}

\begin{figure}
\centering
\includegraphics[scale=0.6]{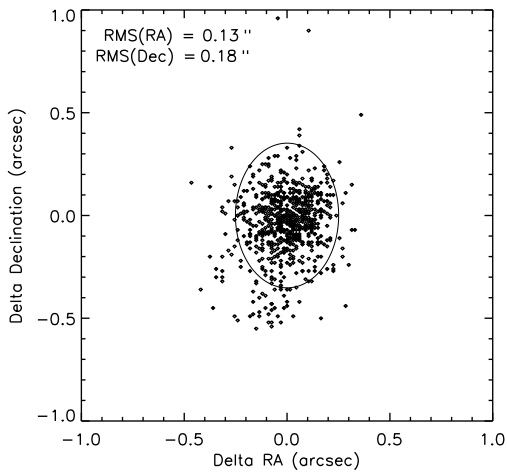}
\caption{A comparison of our source positions with FIRST sources brighter than 10 mJy.  The ellipse shows two times the rms scatter in right ascension and declination.  The values of rms scatter are consistent with that expected due to the accuracy of the FIRST survey.  There does appear to be a slight offset in declination, in the sense that the Stripe 82 positions are biased slightly south of the FIRST positions. }
\label{fig:FIRSToffset10mJy}
\end{figure}

\begin{figure}
\centering
\includegraphics[scale=0.6]{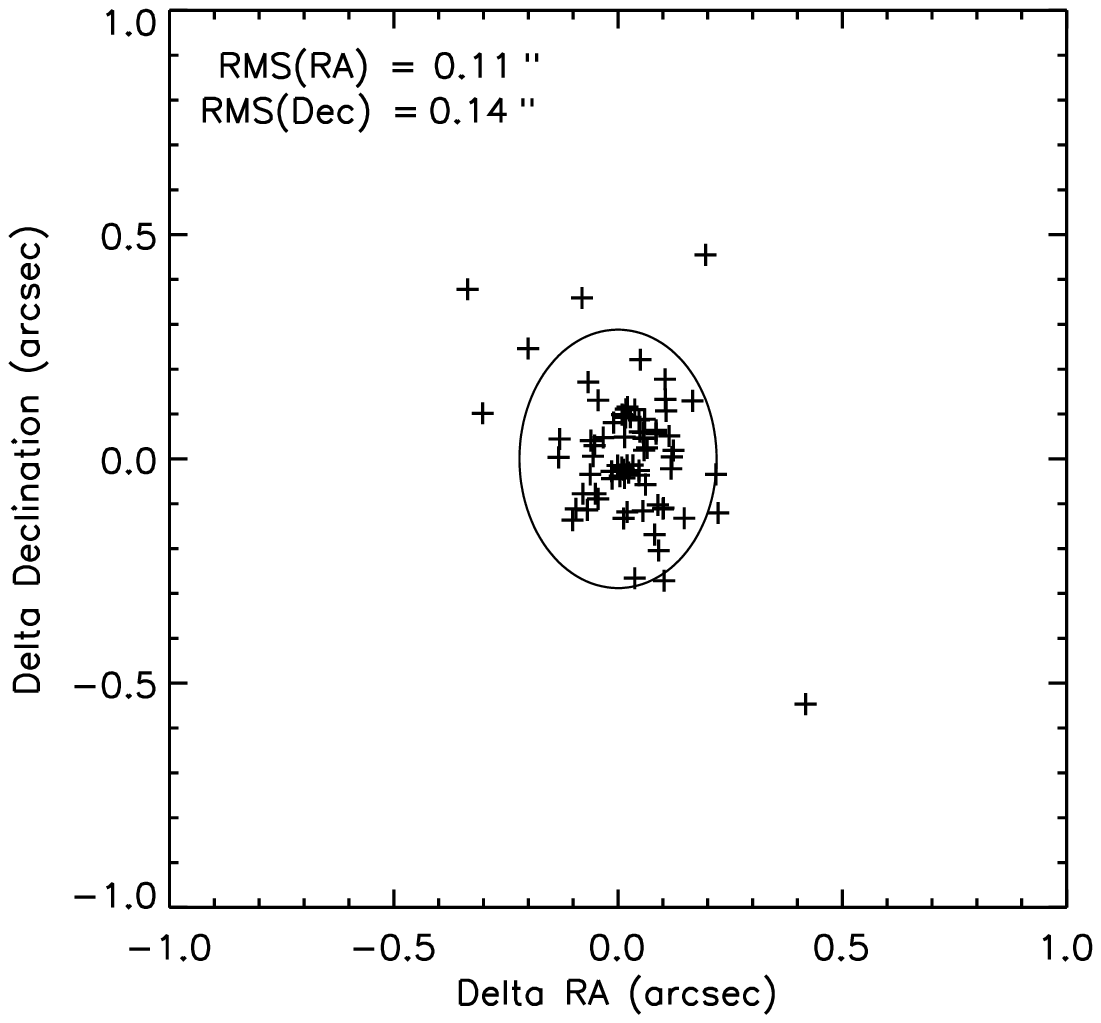}
\caption{A comparison of Stripe 82 source positions with the SDSS DR7 Quasar Catalog \citep{Schneider:2010p2360}.  The comparison was limited to sources brighter than 5 mJy.  The 75 source pairs have rms values of 0.11$^{\prime\prime}$ and 0.14$^{\prime\prime}$ in $\delta$RA and $\delta$Dec, respectively, with mean offsets of 0.02 $\pm$ 0.01, and 0.01 $\pm$ 0.02.  The rms scatter is very good, and there appears to be no offset in right ascension or declination.}
\label{fig:astrometry5mJy}
\end{figure}

\subsection{Astrometry}

To assess the astrometric accuracy of the final maps, we tested the robustness of our catalogued source positions both internally and against external catalogs.  To do the internal test, we utilize the fact that there is a fairly significant amount of overlap between adjacent grid images.  A single source can therefore appear in multiple grid images, and calculating the positional offset between observations for each multiply-observed source is a good test of astrometry for weak sources.  

Figure \ref{fig:posncheck} shows the positional offsets for multiply-observed sources with S $>$ 1 mJy.  Each point is plotted twice - the second time with the sign reversed - since the order is arbitrary.  The ellipse shows two times the rms scatter in right ascension and declination.  Note that we tried restricting the matching to point sources, but we found that the rms does not improve by putting a limit on deconvolved major axis.  We therefore allow all sources in the matching.  This plot shows 4598 pairs of sources and has an rms of 0.19$^{\prime\prime}$ in both right ascension and declination.  
For comparison, the same analysis for FIRST yields rms values of 0.49$^{\prime\prime}$ and 0.46$^{\prime\prime}$ in right ascension and declination \citepalias{Becker:1995p1065}.  This survey thus achieves an astrometric accuracy approximately 2.5 times better than the FIRST survey.  Although the resolution of the A-configuration is three times better than FIRST, we believe that our astrometry may be limited by bandwidth smearing, which was discussed in Section~\ref{bwsmear}.

For the external checks, we compared against the FIRST survey and the SDSS DR7 Quasar Catalog \citep{Schneider:2010p2360}.  (Note that we would have liked to compare against the VLA calibrator catalog, but there are only seven VLA calibrators in the area covered by these observations $-$ not enough for a statistically significant comparison.)  Figure \ref{fig:FIRSToffset10mJy} shows the comparison with the FIRST survey.  As fainter sources will have inherently worse astrometric accuracy, we limit the comparison to sources that are both point sources \textit{and} have flux densities greater than 10 mJy in the FIRST catalog.   The 672 source pairs represented in Figure \ref{fig:FIRSToffset10mJy} have rms values of 0.13$^{\prime\prime}$ and 0.18$^{\prime\prime}$ in $\delta$RA and $\delta$Dec, respectively, with mean offsets of -0.006$^{\prime\prime}$ $\pm$ 0.005$^{\prime\prime}$, and -0.038$^{\prime\prime}$ $\pm$ 0.007$^{\prime\prime}$, one-fifth the rms accuracy of overlapping sources.  The ellipse again shows two times the rms scatter in right ascension and declination.  These values of rms scatter are no better than the FIRST survey, which had lower resolution;  however, this merely reflects the fact that we are comparing against FIRST and therefore are limited by its accuracy.  There does appear to be a slight offset in declination, in the sense that the Stripe 82 positions are biased slightly south of the FIRST positions.  It is unclear what is causing the plume of points outside the ellipse at -0.15, -0.4, though they seem to occur in lines on the sky.  This may be due to a small systematic calibration error in the positions.  

Figure \ref{fig:astrometry5mJy} shows the comparison between the Stripe 82 catalog and the SDSS DR7 Quasar Catalog \citep{Schneider:2010p2360} for sources brighter than 5 mJy in the FIRST catalog.  Such a comparison can be used to tie together the radio and optical reference frames.  The 75 source pairs have rms values of 0.11$^{\prime\prime}$ and 0.14$^{\prime\prime}$ in $\delta$RA and $\delta$Dec, respectively, with mean offsets of 0.02$^{\prime\prime}$ $\pm$ 0.01$^{\prime\prime}$, and 0.01$^{\prime\prime}$ $\pm$ 0.02$^{\prime\prime}$.  Here there appears to be no significant offset in either right ascension or declination.  We conclude that the astrometric accuracy of the Stripe 82 catalog is excellent, as confirmed by both internal checks and comparisons to independent catalogs.

\subsection{Photometry}
\label{photometry}

\begin{figure}
\centering
\includegraphics[scale=0.7]{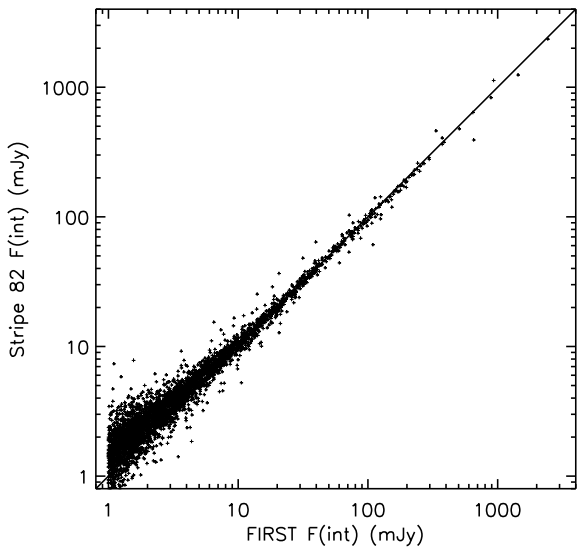}
\caption{A comparison of our integrated flux densities with those from FIRST for sources in both catalogs.  This Figure shows that there is a bias toward higher Stripe 82 values at low flux densities, which is likely caused by the Malmquist bias $-$ see Section~\ref{photometry}.  Also refer to Figure \ref{fig:malmquist}, which examines the difference between flux densities as a function of source size.}
\label{fig:FIRSTfluxcheck}
\end{figure}

We next check the photometric accuracy of the Stripe 82 catalog.  There are multiple sources of photometric error, including resolution bias, CLEAN bias, the model of the primary beam, and bandwidth smearing.  We first compare our integrated flux densities with those from the FIRST survey for sources in both catalogs.  This is shown in Figure \ref{fig:FIRSTfluxcheck}.  We restricted the matching to point sources with FIRST S$_{int}$ $>$ 1 mJy and with only one source per `island', which is a concept used by the source-finding algorithm `HAPPY' and is basically a contiguous set of pixels above a user-defined threshold \citep{White:1997p2118}.  This Figure shows that there is a bias toward higher Stripe 82 values at low flux densities despite the fact that the Stripe 82 observations may resolve out extended flux more so than FIRST.  This offset is likely due to the Malmquist bias.  For sources that are unresolved in the Stripe 82 data, there would not be any bias if the surveys were equally sensitive.  For sources that are resolved in the higher-resolution Stripe 82 data, however, the Stripe 82 data is not as sensitive as the FIRST survey.  A source that is small compared with the FIRST beam ($<$ 3-4$^{\prime\prime}$ FWHM) but is resolved in the Stripe 82 data will have a lower peak flux density in the Stripe 82 data, which can put it below the Stripe 82 detection threshold.  It will appear in the Stripe 82 catalog only if a noise fluctuation causes the apparent peak flux to rise above 1 mJy, in which case it will also appear brighter than the matched FIRST source.  

\begin{figure*}
\centering
\includegraphics[scale=0.66]{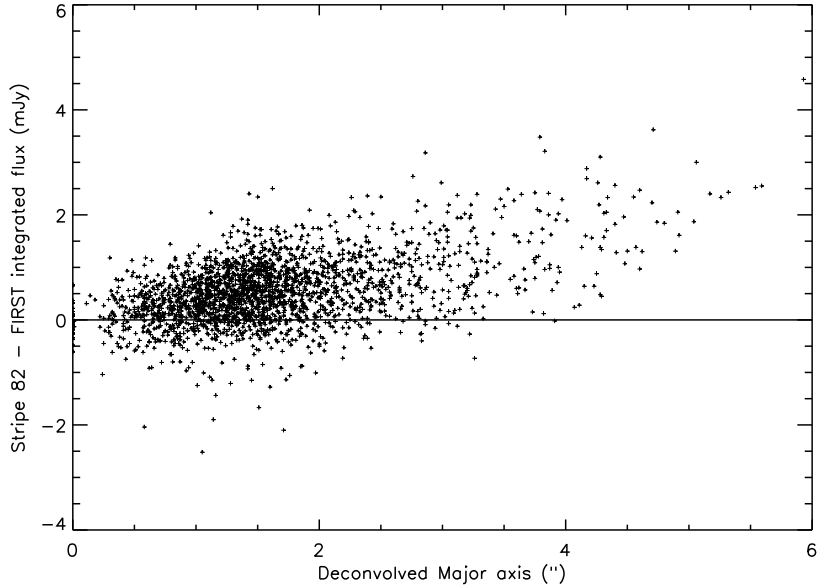}
\caption{The difference between the Stripe 82 and FIRST integrated flux densities as a function of Stripe 82 deconvolved major axis.  As the source size increases, the distribution shifts toward higher Stripe 82 fluxes as the fainter sources disappear, implying that the Malmquist bias is at play.  Note that these are weak sources.}
\label{fig:malmquist}
\end{figure*}

\begin{figure}
\centering
\includegraphics[scale=0.7]{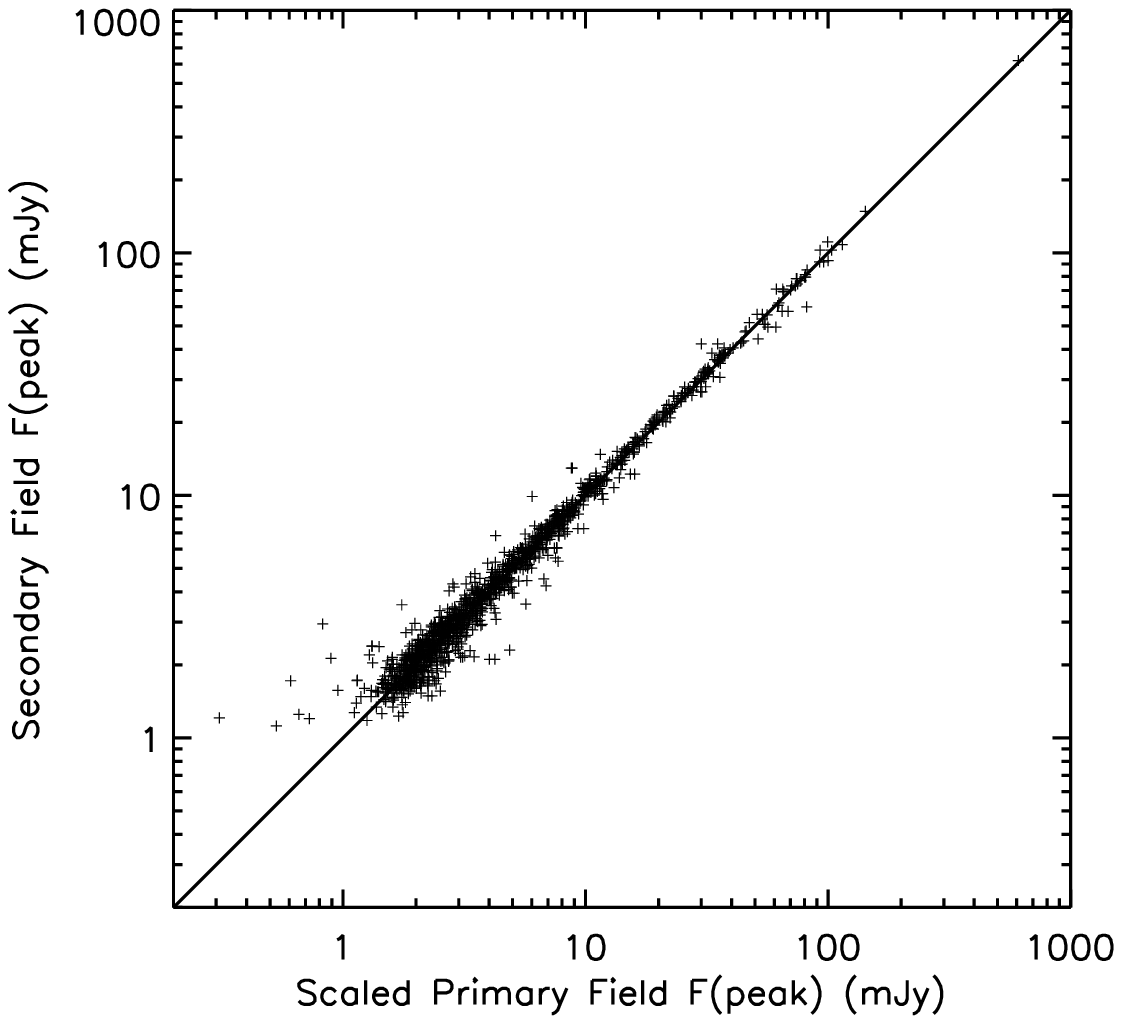}
\caption{A comparison of flux densities for pairs of multiply-observed sources.  The ``primary field" refers to the field where the source is closest to the center, and it includes a correction factor to account for the radially-dependent sensitivity of the primary beam.}
\label{fig:fluxcheck2}
\end{figure}

To test whether this is correct, we have plotted the difference between the Stripe 82 and FIRST integrated flux densities as a function of Stripe 82 deconvolved major axis.  This plot is shown in Figure \ref{fig:malmquist}.  We have limited the plot to weak (S $<$ 3 mJy) sources, where the effect should be more prominent.  We see that for unresolved sources, with a deconvolved major axis around 0.0$^{\prime\prime}$, the distribution of points is fairly symmetric about zero.  As the source size increases, however, the distribution shifts up toward higher Stripe 82 fluxes as the fainter sources disappear.  This confirms that the Malmquist bias is at play.  

Another way to test for errors in the photometry is to again utilize multiply-observed sources.  The idea is that if photometric errors occur in a certain field, then after the primary beam correction, the error will be a function of distance from field center.  We show flux densities for pairs of multiply-observed sources in Figure \ref{fig:fluxcheck2}.  The ``primary field" refers to the field where the source is closest to the center.  To make the fluxes comparable, we have corrected the peak flux of the primary field by the ratio of the primary beam corrections for the two fields.  This is thus also a test of the primary beam model.  Figure \ref{fig:fluxcheck2} shows a very tight correlation over almost three orders of magnitude, indicating that the relative photometric errors are small.  

Outliers in such a plot could also indicate variability, as the observations of the fields were taken anywhere from minutes to weeks apart.  The variability would have to be strong, since each field is the combination of several individual observations and this would tend to damp out short term variability.  There appear to be several prominent outliers in this Figure, all fainter than 1 mJy in the primary field.  Upon closer inspection, however, these sources all show up in fields that are unusually noisy or have wave-like patterns due to incompletely-removed interference.  If we throw out those sources, the width of the remaining correlation still implies we are only sensitive to variability of a factor of several on the faint end.  
We will conduct a search for variability in a follow-up paper (Hodge et al., in prep.).

All VLA snapshot images suffer from a ``CLEAN bias", a photometric defect where the flux density of sources is systematically underestimated \citep{Becker:1995p1065, Condon:1998p1324}.  While our maps were not created from snapshot images, which should preclude a CLEAN bias, we searched for a systematic photometric bias by planting fake sources in the calibrated u-v data and repeating the exact same processing steps on the data to recover the sources.  The planted sources were inserted with the AIPS task 'UVMOD' and consisted of point sources with a range of peak flux densities and distances from field center.  We randomly selected thirteen fields in which to insert the fake sources, and these fields had rms noise values between 45 and 73 $\mu$Jy beam$^{-1}$.  After recovering the planted sources with HAPPY, we compared the measured flux density of each source with its true flux density.  

\begin{figure*}
\centering
\includegraphics[scale=0.5]{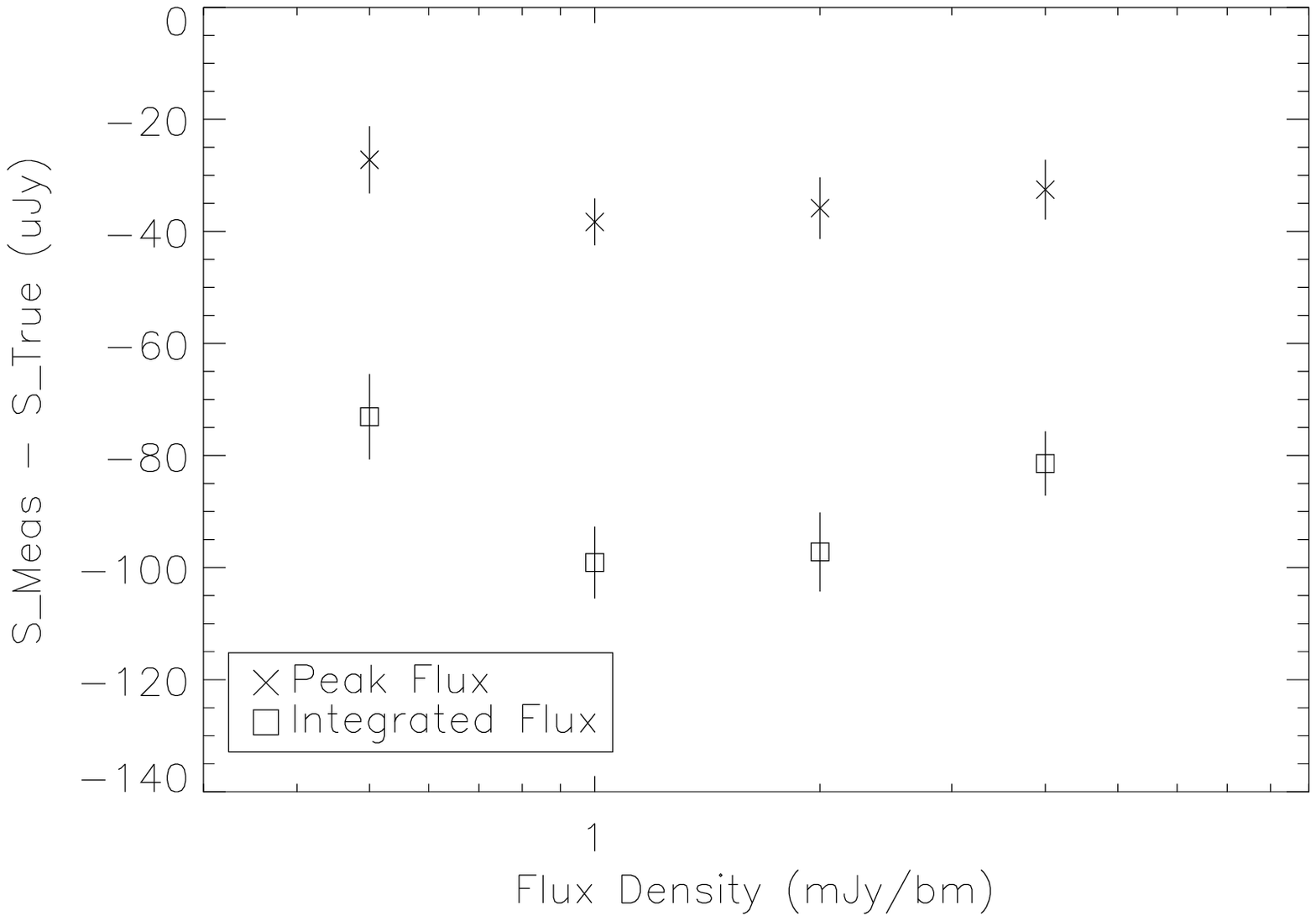}
\hfill
\includegraphics[scale=0.5]{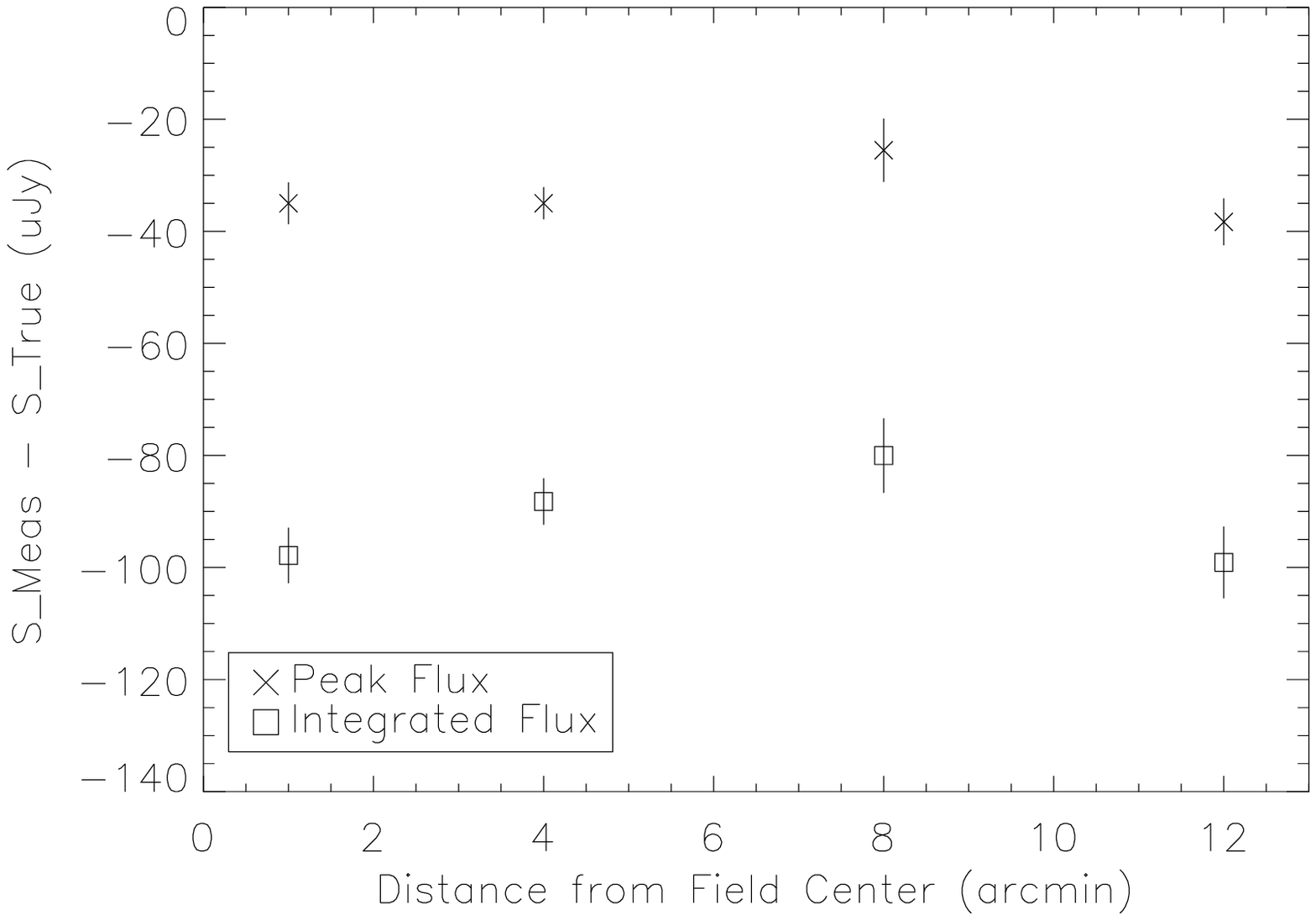}
\caption{Top: The difference between measured and true flux density as a function of artificial source strength.  For each value of source flux density, we inserted twenty sources in a circle of diameter 12$^{\prime}$.  The data points at the top of the plot represent the bias calculated using peak flux, while the data points at the bottom represent the bias calculated using integrated flux.  Bottom: The difference between measured and true flux density as a function of radial distance to the field center.}
\label{fig:Cleanbias}
\end{figure*}

In Figure \ref{fig:Cleanbias} (left), we show the difference between measured and true flux density as a function of planted source strength.  For each value of source flux density, we inserted twenty sources in a circle of diameter 12$^{\prime}$ in the u-v data.  The data points at the top of the plot represent the bias calculated using peak flux, while the data points at the bottom represent the bias calculated using integrated flux.  The bias is clearly small (roughly 35 $\mu$Jy beam$^{-1}$ for peak flux density and sources 12$^{\prime}$ from the beam center) and it does not appear to depend strongly on source strength.  


To investigate the radial dependence of the bias seen in our Stripe 82 images, we inserted fake 1 mJy sources in the u-v data at four different distances from field center.  The result, shown in Figure \ref{fig:Cleanbias} (right), is that the bias does not seem to depend on distance from field center.  At all distances, it maintains a value of roughly 35 $\mu$Jy beam$^{-1}$ as derived from values of peak flux.  This is different than the CLEAN bias, where the functional form describing this effect happened to be such that, once the grid images were corrected for the shape of the primary beam prior to being co-added, the radial dependence disappeared \citep{Becker:1995p1065}.  This complicates matters somewhat, since the correction factor of a particular source would be a function of its position in each grid image that went into the co-added map.  Because of this, and since the bias is relatively small, we do not correct for it in the resultant Stripe 82 catalog.  The result is that the bias for the \textit{grid} sources will be 35 $\mu$Jy beam$^{-1}$ at field center and increase monotonically outward.  Grid sources far from the field center will therefore see larger values of the bias.  Note, however, that when the grid images are co-added, these sources will also contribute less to the co-added source flux density due to weighting by the primary beam.

The largest effect on the photometry is therefore likely due to bandwidth smearing.  The effect of bandwidth smearing is to reduce the peak flux density of the sources while keeping the integrated flux density the same.  Figure \ref{fig:bwbeam6} (Section \ref{bwsmear}) shows the measured major axis size versus distance from the field center for all the grid image sources with S$_{peak}$ $>$ 5 mJy beam$^{-1}$.  This Figure shows that the smearing is relatively small out to 10$^{\prime}$. Using the dimensionless parameter $\beta$ =  $\frac{\delta\nu}{\nu_{\circ}}\frac{\theta_{\circ}}{\theta_{FWHM}}$ as an indicator \citep[see][]{Bridle:1999p371}, source peak flux densities are expected to be reduced by $\le$10\% out to that radius. 
This is an additional source of uncertainty that is not reflected in the generic equation for the uncertainty of catalog flux densities.  


\subsection{Morphology}

\begin{figure*}
\centering
\includegraphics[scale=0.4]{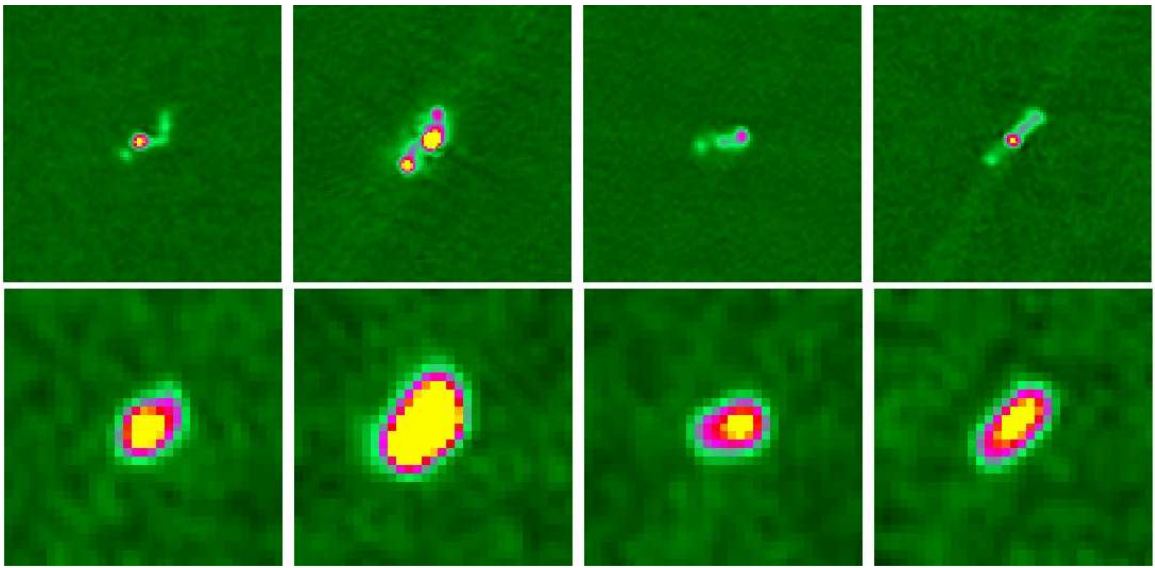}
\caption{The top row shows 1$^{\prime}$ $\times$ 1$^{\prime}$ images resulting from this survey of several known SDSS quasars.  The bottom row shows the corresponding region of sky as seen by the FIRST survey.  The columns show (1) J011129+003431, (2) J015243+002039, (3) J221308+003655, and (4) J222744+005730. }
\label{fig:resolvedQSOs}
\end{figure*}

One of the main differences between this survey and the FIRST survey is the approximately threefold increase in angular resolution.  In addition to the enhanced detail provided on all sources, some sources are now resolved into multiple sources.  To estimate the effect of this increase, we measured how many FIRST sources were resolved into multiple components in our Stripe 82 radio catalog by matching the catalogs with a matching radius of 5$^{\prime\prime}$.  We found that 1302 out of the 9328 detected FIRST sources were resolved into multiple components, or roughly 14\%.  Some examples of such sources are shown in Figure \ref{fig:resolvedQSOs} as 1$^{\prime}$ $\times$ 1$^{\prime}$ postage stamps.  The images in the top row are the images from this survey, while the bottom row shows the corresponding cutouts from the FIRST survey.  These sources all correspond to known SDSS quasars. 

\begin{figure}
\centering
\includegraphics[scale=0.48]{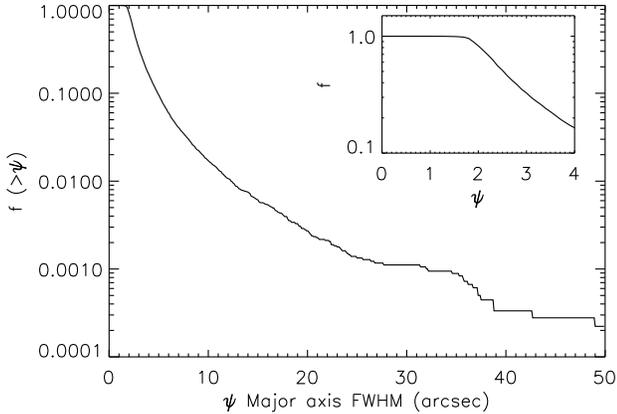}
\caption{The cumulative distribution function of major axis FWHM values for all Stripe 82 sources.  The inset highlights small values of the major axis up to $\psi$ $=$ 4$^{\prime\prime}$.}
\label{fig:fmajcdf}
\end{figure}

The cumulative distribution function of major axis FWHM values for all sources in the catalog is shown in Figure \ref{fig:fmajcdf}.  
The inset shows the function for small values of the major axis ($\psi$ $<$ 4$^{\prime\prime}$).  The high fractions achieved for these small values illustrate that 
over 17,300 sources, or 96\% of the catalog, are larger than the CLEAN beam size of 1.8$^{\prime\prime}$.  This does not represent the fraction of resolved sources, as we established in Section \ref{bwsmear} that the effective beam size is actually larger due to bandwidth smearing.  We use 2.3$^{\prime\prime}$ as an estimate of the major axis value below which the vast majority of smeared beams lie (see Figure \ref{fig:bwbeam3}).  Sources at the edge of the coverage will have smeared beams larger than this value, but many sources will also have smeared beams somewhat smaller.  Using this value, we find that $\sim$60\% of the catalog sources have been resolved.  
For comparison's sake, only $\sim$20\% of the sources in the FIRST catalog are larger than the FIRST CLEAN beam of 5.4$^{\prime\prime}$.  
Our value is comparable to other 1.4 GHz surveys of similar resolution, which probe much fainter flux densities (rms of a few $\mu$Jy) and find that $\sim$70\% of their sources are resolved when they adopt a classification method utilizing the results of JMFIT \citep{Owen:2008p136, Miller:2008p114, Schinnerer:2010p384}.

\section{COMPLETENESS}
\label{completeness}

The are several factors that affect the completeness of a source catalog.  The most obvious obstacle to catalog completeness in our case is the fact that the rms sensitivity varies over the area covered, and sometimes quite dramatically (Figure \ref{fig:coveragemap} and Section~\ref{depth}).  
In addition to varying rms sensitivity, there are several additional factors that also influence the completeness, including CLEAN bias, bandwidth smearing, and the resolution bias.  
Here, we will compare the Stripe 82 catalog to existing catalogs in order to estimate the completeness of the survey.

In Section \ref{resbias}, we matched the Stripe 82 catalog to the FIRST survey catalog and found that 22\% of FIRST sources have no counterpart in the Stripe 82 catalog.  As extended sources are more likely to be resolved out when they are faint, the completeness increases with decreasing source size and increasing source strength.  To further understand the FIRST sources missing from the Stripe 82 catalog, we compared their optical match rate with the match rate of FIRST sources that were recovered in the Stripe 82 catalog.  For the optical data we used the deep SDSS imaging of Stripe 82, which is available through CasJobs\footnote{Specify Context $=$ ``Stripe 82" and runs 106 or 206 to retrieve the deeper co-added photometry.}.  Cross-matching the catalogs with a matching radius of 5$^{\prime\prime}$ reveals that FIRST sources with Stripe 82 radio counterpart(s) have an optical match rate of 72\%, 
whereas FIRST sources without Stripe 82 radio counterparts have an optical match rate of only 46\%.  
The background rate is unimportant, as it would be the same for both scenarios.  This drop in match rate therefore indicates that some of the undetected FIRST ``sources" may not have been sources at all, but noise peaks and image artifacts.  Based on the percentage drop, it is possible that as many as 26\% of the unmatched FIRST sources may have been false detections.  We checked the sky distribution of unmatched sources, and some of them do group into narrow lines, a pattern indicative of sidelobes.   However, after examining some of the unmatched sources by eye, we conclude that the percentage of false sources is probably not quite this high.  While some of the unmatched sources are indeed sidelobes, the vast majority are undetected for other reasons, typically because they have been completely resolved out, or are visible but just below the Stripe 82 detection threshold.  Most are faint sources right at the FIRST detection threshold of 1 mJy.

We next looked to the population of previously-known radio quasars to see how many were recovered in the Stripe 82 catalog.  This population should provide a sample of sources where (1) the objects are not expected to be significantly extended, and (2) the objects are confirmed as real sources.  There are 3885 quasars from the SDSS DR7 Quasar Catalog \citep{Schneider:2010p2360} in the area covered by our observations.  The FIRST catalog has sources matching 229 out of the 3885 
within a matching radius of 5$^{\prime\prime}$.  Of those 229, the Stripe 82 radio catalog recovers 223 of them.  Of the 6 missing, all are visible by eye in the cutout server at the position of the FIRST sources.  Five of them appear to be barely resolved out, with faint peak flux densities just below the catalog threshold at their position.  The remaining source is brighter, with a peak flux density approaching 1 mJy, but it has been excluded from the catalog due to a bright source in the vicinity, requiring the rms threshold to be automatically raised to avoid adding spurious sidelobes to the source catalog.   The new catalog is therefore highly complete for the radio quasar population, recovering over 97\% of the FIRST-detected quasars.  

The discussion so far has been limited to a comparison with the FIRST survey, which has a detection limit of only 1 mJy.  To estimate the Stripe 82 survey completeness below 1 mJy, we will compare our 1.4 GHz source counts with those from the VLA-COSMOS survey \citep{Bondi:2008p1129}, which was also conducted at high resolution (1.5"), but achieves a much lower rms sensitivity of 11 $\mu$Jy.  We calculated the Stripe 82  differential counts by dividing the number of sources in each flux density bin by the survey area of that bin and the bin width.  We normalized the differential counts using the historical value of $S^{-2.5}$.  We used the peak flux density for point sources and the integrated flux density for extended sources, and we defined a source as extended if it had a ratio of integrated to peak flux density $S_{int}/S_{peak}$ $>$ 1.1.  To account for the varying sensitivity of the survey, we also include a correction for the effective survey area.  The effective area was calculated by weighting each source in the bin by the inverse of the area over which the source could have been detected \citep{Katgert:1973p171, White:1997p2118}, which we found by summing the area in the coverage map with rms $\le$ S$_{peak}$ $/$ 5.  

\begin{figure}
\centering
\includegraphics[scale=0.48]{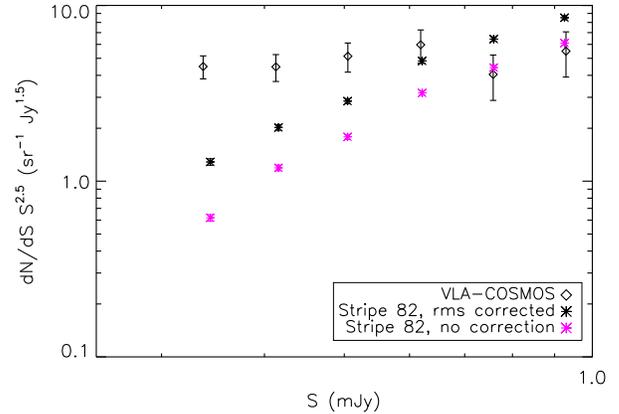}
\caption{Stripe 82 source counts below 1 mJy compared to the source counts from the VLA-COSMOS survey \citep{Bondi:2008p1129}.  The counts are shown with and without the correction for varying survey sensitivity.   }
\label{fig:lognlogs}
\end{figure}

The Stripe 82 source counts up to $\sim$1 mJy are shown in Figure \ref{fig:lognlogs} compared to the source counts from the VLA-COSMOS survey \citep{Bondi:2008p1129}.  The Stripe 82 counts are shown with and without the correction for varying survey sensitivity.  The VLA-COSMOS source counts from \citet{Bondi:2008p1129} also include this correction, as well as empirical corrections for bandwidth smearing, noise bias, fitting errors, and resolution bias. The difference between the VLA-COSMOS counts and the Stripe 82 counts therefore indicates the remaining incompleteness in the Stripe 82 catalog.  

The Stripe 82 survey is clearly incomplete for sources near the detection limit.  The source counts are a factor of 3.5 lower than the VLA-COSMOS source counts for the lowest flux density bin ($\langle$S$\rangle$ $=$ 0.34 mJy).  The discrepancy decreases with increasing flux density, with the Stripe 82 source counts coming into agreement with VLA-COSMOS around S $=$ 0.6 mJy.  The drop-off in source counts for low flux density bins is most likely due primarily to the resolution bias, and catalog users should be aware that the catalog is incomplete for faint and extended sources.

\section{OPTICAL IDENTIFICATION}
\label{optical IDs}

Optical identifications for the Stripe 82 radio sources were determined by cross-matching the radio catalog with the deep SDSS imaging of Stripe 82 introduced in Section~\ref{completeness}.  As this coverage does not extend quite as far in Declination from $\delta = 0$ as our radio coverage, for the purposes of this section we trim our catalog so as not to falsely underestimate the optical matching rate.

To determine a suitable matching radius between the two catalogs, we first examined the matching completeness and efficiency as a function of matching radius.  For this exercise, we used a liberal matching radius of 15$^{\prime\prime}$ and kept the closest SDSS match to each radio source within this radius.  One-to-one matching was required, with SDSS sources flagged once they had been matched once.  To model the underlying background of false matches, we shifted the radio source positions by 1$^{\prime}$ in four different directions and repeated the matching, averaging the results.  

\begin{figure}
\centering
\includegraphics[scale=0.48]{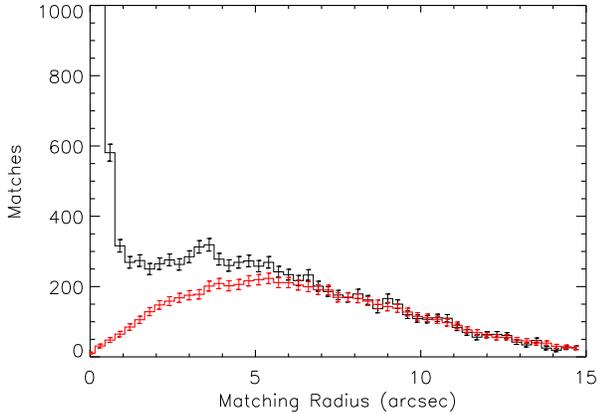}
\caption{Matching our radio catalog to the deep SDSS imaging of Stripe 82: Number of matches versus matching radius.  The black line shows the total number of matches, and the red line shows the background matches calculated by shifting the radio catalog by 1$^{\prime}$ in four directions and repeating the matching, averaging the results.  }
\label{fig:SDSSmatchradius}
\end{figure}

Figure \ref{fig:SDSSmatchradius} shows the matches as a function of matching radius in bins of 0.3$^{\prime\prime}$. 
 The total matches are shown by the black line, and the background matches are shown by the red line.  The total matches peak at N $=$ 4816 for r $<$ 0.3$^{\prime\prime}$, but the y-axis cuts off at N $=$ 1000 in order to show the background more clearly. The difference between the total and background matches is then an estimate of the number of matches that are actual physical associations.  Both the total and background matches show a turnover and eventual decrease in matches with increasing matching radius, a result of the one-to-one matching constraint.  Modeling of the chance background rate in the way described above is not completely accurate, as real matches cannot also be false matches.  Simply shifting one of the catalogs therefore over-predicts the false match rate.  \citet{McMahon:2002p2009} discuss this effect in more detail and derive an expression for the effective background source density.  We have applied a first-order approximation of this correction to the background matches shown in Figure \ref{fig:SDSSmatchradius}, where the true background is then given by:
 \begin{equation}
 N_{\rm true} = N_{\rm obs} \left ( \frac{\sum_{0}^{r}{\rm matches}-\sum_{0}^{r}{\rm background}}{N_{\rm obj}} \right )
 \end{equation}
 where N$_{\rm obj}$ are the number of objects in the radio catalog and the summations go up to some radius, r, where the background is not yet dominating the total matches.  We use a radius of r $=$ 1$^{\prime\prime}$ to calculate the correction, and the corrected background and total matches show good agreement at large values of the matching radius, validating the correction.  

The total matches show a noticeable surplus over the background matches at separations beyond that where the clear peak in total matches has dropped off.  This excess, which persists out to separations of 5$^{\prime\prime}$ and beyond, is not inconsistent with that found by \citet{McMahon:2002p2009} for matching between the FIRST radio catalog and the SERC Automated Plate Measuring Machine (APM) scans of the Palomar Observatory Sky Survey (POSS-I) plates, and it likely has two origins.  The first is that, as radio galaxies often reside in galaxy clusters, by allowing matching separations out to 15$^{\prime\prime}$, we could be matching radio sources to unassociated but nearby cluster members.  The true background is therefore underestimated by simply shifting one of the catalogs and re-matching, as the actual source density in these instances is much higher. The second possibility is that components of multi-component radio sources are being matched to an optical counterpart which does not align with any of the radio components.  In this case the association is real in the sense that the optical and radio sources are related, but without redshift information or a smarter algorithm to definitively tie together the multiple radio components based on morphology, the match rate of the radio sources will be artificially low.

\begin{figure}
\centering
\includegraphics[scale=0.48]{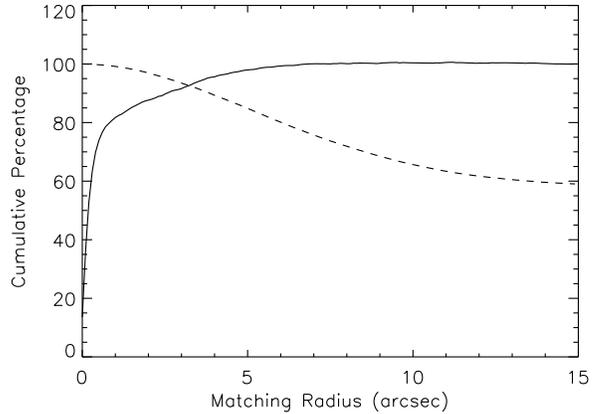}
\caption{The completeness (solid line) and reliability (dashed line) of matching the Stripe 82 radio catalog to the deep SDSS imaging as a function of matching radius.}
\label{fig:SDSScompleteness}
\end{figure}

Choosing a matching radius always comes down to a tradeoff between completeness and reliability, with the decision ultimately informed by what sort of science one hopes to accomplish.  The completeness and reliability as a function of matching radius are shown in Figure \ref{fig:SDSScompleteness}.  
Completeness (solid line) gives the percentage of the physical matches recovered at that matching radius, while reliability (dashed line) gives the percentage of total matches that are likely to be physical.  The surplus in total matches out to fairly large ($\sim$5$^{\prime\prime}$) radii is clearly evident.  The completeness is approaching 98\% at that radius, and the reliability is 85\%.  On the other hand, the matching is extremely reliable for small matching radii.  For separations less than 1$^{\prime\prime}$, 99\% of the matches are likely to be physical, while 97\% are physical for separations less than 2$^{\prime\prime}$.  The values of completeness for 1$^{\prime\prime}$ and 2$^{\prime\prime}$ are 82\% and 88\%, respectively.   

Since we wish to provide a reliable estimate of the optical identifications for the sources in our catalog, we will opt for a small matching radius in spite of the low completeness.  For this matching, we will therefore use a matching radius of 1$^{\prime\prime}$.   Using this radius and an area common to both catalogs (see above), we find a total optical match rate of 44.4\%.  For comparison, the Australia Telescope Hubble Deep Field-South (ATHDF-S) survey \citep{Huynh:2008p2470} reported an optical match rate of 67\% when they compared their radio catalog to observations of the HDF-S by the Cerro Tololo Inter-American Observatory (CTIO).  This imaging data is similar in depth to the co-added SDSS imaging (I $=$ 23.5) used here, but the radio catalog has a detection limit of $\sim$50 $\mu$Jy.  They also found that the I-band magnitude distribution peaked between I $=$ 19 and I $=$ 20 and decreased for higher magnitudes.  They argued that this was a property of the radio sources themselves and not due to the incompleteness in the optical imaging, since the fall-off occurred well above the limiting magnitude of 23.5 in I.  This implies that, while there is surely a significant population of radio sources with optical counterparts too faint to be detected here, the majority are not lurking just below the limiting magnitude.


One situation where one would not necessarily expect there to be an optical counterpart for each radio component is in multiple-component systems such as double-lobed radio galaxies.  To find out if the unmatched population are more likely to be members of such systems than sources with optical counterparts, we perform a search around each unmatched source for nearby neighbors within the Stripe 82 radio catalog.  We find that $\sim$70\% of the unmatched sources have another Stripe 82 radio source within 30$^{\prime\prime}$.  
We then performed the same search around all of the Stripe 82 radio sources, coming up with 80\%.  
We therefore cannot conclude that the unmatched sources are unmatched because they are more likely to be part of multiple-component systems.  They are actually slightly {\it less} likely to have a nearby neighbor, although the difference is small enough that it may not be significant.


The optical counterparts can be further classified into either galaxies or stellar sources based on SDSS photometry.  ``Stellar sources" refer to sources that are not extended.  While a small number of stars emit in the radio, the majority of these radio sources are more likely quasars.  The results break down into: 40.6\% galaxies, 3.8\% stellar, and 55.6\% unmatched.  

\begin{figure}
\centering
\includegraphics[scale=0.48]{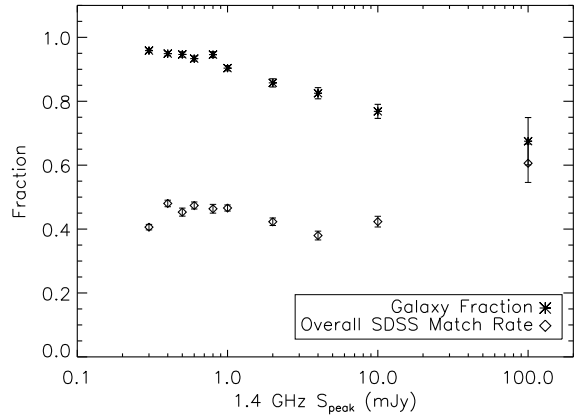}
\caption{Fraction of Stripe 82 radio sources versus 1.4 GHz peak flux density in mJy.  The diamonds give the fraction of radio sources that have an SDSS match within 1$^{\prime\prime}$.  The crosses give the fraction of the matched radio sources that have a galaxy as their optical counterpart based on SDSS photometry.  The remaining sources match to stellar counterparts, the majority of which are quasars.}
\label{fig:SDSSmatchrate}
\end{figure}

Figure \ref{fig:SDSSmatchrate} shows the optical identification of all sources in the radio catalog broken up based on radio source strength.  
Shown in this Figure are two separate quantities: the overall SDSS match fraction, and the fraction of {\it those} objects that correspond to galaxies.  The radio sources have been binned by peak flux density, with the data points marking the beginning of each bin.  The overall SDSS match fraction, designated by the diamond points, shows a slight decrease with decreasing source strength, from $\sim$60\% for sources over 100 mJy to just over 30\% for sources between 0.3 and 0.4 mJy.  The roll-off at the faint end could be a real effect, but it is most likely due to an increasing population of spurious sources at fainter flux densities.  However, the fraction of matches increase from about 4 mJy to 0.4 mJy, showing that the Stripe 82 catalog is very reliable down to 0.4 mJy.  

Meanwhile, the galaxy fraction, designated by the crosses, shows a smooth increase with decreasing source strength, from $\sim$65\% for galaxies at the bright end to $\sim$95\% at the faint end.  This is to be expected, as it is well-known that the source population sampled by bright radio surveys is primarily composed of quasars and powerful radio galaxies, while fainter surveys sampling the millijansky level and below are increasingly dominated by a mixture of low-luminosity radio-quiet AGN and star-forming galaxies \citep[for e.g.,][]{Afonso:2006p1216, Smolcic:2008p14, Seymour:2008p1695, Ibar:2010p281}.  Nevertheless, we do expect to detect previously-unknown radio quasars with this survey. To see how many new quasars the survey detects, we once again turn to a comparison with the SDSS DR5 QSO catalog.  

\renewcommand{\arraystretch}{0.8}		
\begin{table}
\centering
  \caption{Newly Detected Radio Quasars}  \label{tab:NEWqso} 
\scalebox{0.78}
 {
\begin{tabular}{llrrrrcc}
   \hline\hline \\ [-2ex]
      \multicolumn{1}{c}{\textbf{RA$^{a}$}} & \multicolumn{1}{c}{\textbf{Dec}} & \multicolumn{1}{c}{\textbf{S$_{\rm pk}$$^{b}$}} & \multicolumn{1}{c}{\textbf{S$_{\rm int}$$^{c}$}} & \multicolumn{1}{c}{\textbf{Maj$^{d}$}} & \multicolumn{1}{c}{\textbf{Min$^{e}$}} & \multicolumn{1}{c}{\boldmath{$z$}$^{f}$} & \multicolumn{1}{c}{\boldmath{$m_{\rm r}$}$^{g}$} \\
       & & \multicolumn{1}{c}{\textbf{(mJy)}} & \multicolumn{1}{c}{\textbf{(mJy)}} & \multicolumn{1}{c}{\textbf{($^{\prime\prime}$)}} & \multicolumn{1}{c}{\textbf{($^{\prime\prime}$)}} &  & \multicolumn{1}{c}{\textbf{(mag)}} \\     [0.5ex] 
       \hline
       \\  [-1.8ex]
    00 39 28.059  & +00 09 53.06 &    0.53 &    0.97 &  1.67 &  1.60 &  1.010 &  18.23 \\
    00 42 13.471  & +00 59 17.47 &    0.51 &    0.64 &  1.52 &  0.00 &  0.329 &  18.58 \\
  00 43 23.433  & $-$00 15 52.37 &   13.99 &   15.42 &  0.60 &  0.55 &  2.798 &  18.47 \\
    00 48 49.630  & +01 05 43.76 &    0.78 &    1.52 &  1.76 &  1.74 &  2.614 &  19.00 \\
    00 53 46.745  & +00 53 43.23 &    0.34 &    0.33 &  0.86 &  0.00 &  2.346 &  19.71 \\
  01 00 47.728  & $-$00 37 51.73 &    0.34 &    0.61 &  1.64 &  1.58 &  0.733 &  19.27 \\
  01 01 13.382  & $-$00 29 48.23 &    0.79 &    1.28 &  1.91 &  0.85 &  1.336 &  19.61 \\
    01 03 29.444  & +00 40 53.47 &    3.57 &   32.94 &  7.71 &  3.31 &  1.433 &  18.39 \\
  01 05 45.633  & $-$00 54 48.58 &    0.47 &    0.87 &  1.69 &  1.66 &  2.011 &  20.34 \\
    01 06 19.245  & +00 48 23.55 &    0.41 &    0.38 &  0.00 &  0.00 &  4.449 &  19.03 \\
  01 15 36.870  & $-$00 00 09.93 &    0.46 &    0.81 &  1.99 &  1.18 &  0.532 &  20.68 \\
    01 17 58.822  & +00 20 21.48 &    0.34 &    0.30 &  0.00 &  0.00 &  0.613 &  17.89 \\
  01 21 36.736  & $-$00 18 12.76 &    7.70 &   32.80 &  5.66 &  1.47 &  0.764 &  19.08 \\
  01 27 50.413  & $-$00 09 19.52 &    0.53 &    0.60 &  1.08 &  0.00 &  0.437 &  19.98 \\
    01 28 24.219  & +00 19 25.16 &    0.46 &    0.44 &  0.00 &  0.00 &  0.420 &  19.55 \\
  01 36 56.322  & $-$00 46 23.74 &    0.48 &    0.45 &  0.82 &  0.00 &  1.716 &  18.16 \\
  01 37 22.496  & $-$00 35 59.96 &    0.38 &    0.33 &  1.01 &  0.00 &  1.815 &  19.01 \\
  01 37 22.985  & $-$00 17 46.43 &    0.48 &    0.49 &  0.38 &  0.29 &  1.442 &  18.76 \\
  01 39 18.267  & $-$01 02 02.26 &    0.62 &    0.69 &  1.30 &  0.00 &  1.478 &  17.94 \\
    01 39 29.536  & +00 13 27.44 &    0.46 &    0.94 &  2.56 &  1.14 &  2.103 &  18.78 \\
    01 40 23.770  & +01 00 43.45 &    0.45 &    1.04 &  2.09 &  2.01 &  1.416 &  20.10 \\
    01 40 40.704  & +00 17 58.18 &    0.62 &    0.64 &  0.42 &  0.27 &  0.404 &  19.19 \\
    01 41 03.818  & +01 05 01.00 &    0.72 &    2.57 &  3.13 &  2.64 &  0.789 &  19.42 \\
    01 45 52.554  & +00 51 51.70 &    0.55 &    1.04 &  2.32 &  1.10 &  0.906 &  19.75 \\
    01 46 41.196  & +01 08 15.92 &    0.37 &    0.42 &  1.49 &  0.00 &  0.980 &  17.27 \\
  01 49 05.268  & $-$01 14 04.76 &    0.56 &    0.66 &  1.30 &  0.00 &  2.100 &  19.70 \\
  01 49 50.957  & $-$01 03 14.06 &    0.68 &    0.70 &  0.70 &  0.00 &  1.074 &  17.52 \\
  01 49 58.556  & $-$00 30 23.28 &    0.38 &    0.83 &  2.08 &  1.81 &  2.114 &  20.00 \\
  01 50 34.542  & $-$00 02 00.79 &    0.36 &    0.49 &  1.47 &  0.66 &  1.739 &  19.68 \\
  01 50 57.524  & $-$00 45 47.39 &    0.99 &    1.11 &  0.74 &  0.51 &  3.056 &  20.46 \\
  01 51 34.818  & $-$00 00 10.44 &    0.48 &    1.15 &  2.15 &  2.07 &  0.346 &  19.36 \\
  01 52 13.365  & $-$00 43 02.44 &    0.34 &    1.00 &  3.20 &  1.83 &  1.025 &  19.00 \\
    01 52 49.766  & +00 23 14.53 &    0.80 &    0.78 &  0.00 &  0.00 &  0.590 &  18.02 \\
  01 55 28.644  & $-$00 38 56.57 &    0.35 &    0.36 &  0.32 &  0.00 &  1.382 &  19.51 \\
  01 55 53.845  & $-$01 08 22.32 &    0.90 &    0.83 &  0.00 &  0.00 &  1.315 &  19.53 \\
    01 56 50.285  & +00 53 08.54 &    0.89 &    0.81 &  0.00 &  0.00 &  1.651 &  18.64 \\
  01 57 33.879  & $-$00 48 24.66 &    0.50 &    0.53 &  0.52 &  0.35 &  1.545 &  18.41 \\
  01 58 19.767  & $-$00 12 22.05 &    0.79 &    0.91 &  1.18 &  0.00 &  3.302 &  18.67 \\
    01 59 21.138  & +00 20 14.64 &    0.65 &    0.78 &  1.11 &  0.35 &  0.939 &  19.89 \\
  02 01 22.776  & $-$00 02 48.62 &    0.48 &    0.57 &  0.77 &  0.75 &  1.509 &  19.48 \\
  02 02 58.940  & $-$00 28 07.08 &    0.35 &    0.49 &  1.49 &  0.75 &  0.339 &  19.29 \\
  02 08 18.643  & $-$00 05 33.76 &    0.34 &    0.36 &  0.44 &  0.42 &  0.493 &  19.84 \\
  02 11 01.007  & $-$00 44 01.95 &    0.35 &    1.10 &  2.68 &  2.57 &  1.362 &  18.64 \\
    02 13 37.624  & +00 25 56.61 &    0.53 &    0.60 &  0.77 &  0.57 &  1.644 &  19.17 \\
  02 14 47.016  & $-$00 32 50.66 &    0.57 &    0.57 &  0.47 &  0.00 &  0.349 &  19.11 \\
    02 17 35.439  & +00 14 56.28 &    0.70 &    1.26 &  1.67 &  1.53 &  1.170 &  20.30 \\
    22 07 59.418  & +00 17 22.51 &    0.39 &    0.46 &  0.79 &  0.74 &  0.367 &  18.77 \\
  22 08 50.458  & $-$00 02 34.03 &    0.53 &    1.58 &  2.81 &  2.28 &  1.138 &  20.16 \\
  22 08 51.965  & $-$01 06 03.47 &    0.46 &    0.50 &  0.69 &  0.37 &  0.352 &  18.10 \\
  22 13 37.968  & $-$00 43 05.90 &    0.44 &    0.58 &  1.81 &  0.00 &  1.182 &  17.47 \\
    22 19 10.546  & +00 56 06.45 &    0.36 &    0.55 &  2.10 &  0.22 &  1.194 &  20.07 \\
  22 20 30.090  & $-$00 22 16.29 &    1.43 &    1.50 &  0.66 &  0.00 &  0.602 &  20.96 \\
    22 24 05.954  & +00 18 29.18 &    0.36 &    0.42 &  1.69 &  0.00 &  0.457 &  19.17 \\
    22 25 18.681  & +01 12 43.14 &    0.41 &    1.00 &  2.31 &  1.99 &  0.700 &  20.53 \\
    22 25 32.862  & +00 57 14.47 &    0.54 &    0.61 &  1.16 &  0.00 &  0.625 &  19.90 \\
  22 26 06.184  & $-$00 32 32.46 &    0.41 &    0.22 &  0.00 &  0.00 &  0.579 &  20.60 \\
  22 26 39.188  & $-$00 03 42.61 &    0.37 &    0.29 &  1.21 &  0.00 &  1.233 &  19.58 \\
  22 27 26.285  & $-$00 28 55.49 &    0.33 &    0.39 &  0.74 &  0.69 &  0.843 &  17.96 \\
  22 31 46.180  & $-$00 27 45.75 &    0.60 &    1.13 &  1.77 &  1.62 &  1.534 &  19.45 \\
    22 34 38.535  & +00 57 30.00 &    0.32 &    0.86 &  3.44 &  1.32 &  2.817 &  19.19 \\
      22 37 04.932  & $-$01 05 51.51 &    0.59 &    0.70 &  1.11 &  0.27 &  1.011 &  18.90 \\
    22 39 25.934  & +00 03 40.37 &    0.61 &    0.59 &  0.56 &  0.00 &  0.586 &  19.63 \\
    22 42 56.471  & +00 51 55.13 &    0.52 &    1.41 &  3.25 &  1.53 &  0.420 &  20.02 \\
    22 45 48.917  & +00 07 19.94 &    0.36 &    1.58 &  3.50 &  3.14 &  0.943 &  19.52 \\
    22 49 33.044  & +01 04 55.79 &    0.78 &    0.92 &  1.04 &  0.34 &  0.264 &  18.83 \\
  23 01 31.798  & $-$00 07 16.37 &    0.41 &    0.70 &  1.51 &  1.48 &  2.103 &  19.09 \\
  23 07 28.908  & $-$01 16 09.04 &    1.84 &    2.47 &  1.51 &  0.41 &  1.971 &   0.00 \\
    23 12 10.762  & +00 27 28.02 &    0.34 &    0.60 &  1.59 &  1.54 &  1.363 &  20.08 \\
    23 15 10.083  & +00 56 05.57 &    0.33 &    0.43 &  1.92 &  0.00 &  1.438 &  20.20 \\
  23 15 40.295  & $-$00 55 28.95 &    0.30 &    0.52 &  1.65 &  1.50 &  0.933 &  19.50 \\
  23 16 45.044  & $-$00 11 29.54 &    0.37 &    3.23 &  5.02 &  4.92 &  0.596 &  20.44 \\
  23 17 11.791  & $-$00 36 03.91 &    0.32 &    0.78 &  2.30 &  2.01 &  0.186 &  17.82 \\
    23 17 42.594  & +00 05 34.84 &    0.35 &    0.70 &  1.85 &  1.80 &  0.321 &  17.90 \\
  23 19 58.690  & $-$00 24 49.19 &    0.43 &    1.99 &  4.26 &  2.70 &  1.891 &  18.13 \\
  23 20 08.944  & $-$00 46 17.47 &    0.48 &    1.32 &  3.12 &  1.70 &  1.257 &  19.82 \\
  23 20 20.728  & $-$00 20 10.87 &    0.46 &    2.90 &  4.80 &  3.52 &  1.158 &  19.91 \\  
 \\ [-1.8ex] \hline\hline \\[-1.9ex]
  \end{tabular}
 }
  \footnotetext{NOTES. $-$ $^{a}$Radio source position (J2000).  $^{b}$1.4 GHz peak flux density (mJy).  $^{c}$1.4 GHz integrated flux density (mJy).  $^{d}$Deconvolved major axis.  $^{e}$Deconvolved minor axis.  $^{f}$Redshift from the DR7 quasar catalog \citep{Schneider:2010p2360}.  $^{g}$Apparent magnitude in the r-band from the same catalog.} 
 \end{table}
\renewcommand{\arraystretch}{1.0}

Recall that the FIRST catalog matches to 229 out of 3885 quasars from the SDSS DR7 Quasar Catalog \citep{Schneider:2010p2360} within a matching radius of 5$^{\prime\prime}$.  
Of those, the Stripe 82 catalog recovers 223.  In addition to those 223, the new catalog also has radio sources matching 76 quasars not previously detected by FIRST.  
The total fraction of spectroscopic quasars that are radio sources to the depth probed here is therefore 7.7\% ($\pm$ 0.4\%).  Conversely, the fraction of radio sources that match to spectroscopic quasars is 1.7\% ($\pm$ 0.2\%).  
To determine the reliability of these matches, we applied the same analysis as above to characterize a match between our catalog and the SDSS DR7 Quasar Catalog.  While the statistics were not adequate to produce detailed plots of reliability and completeness as a function of matching radius, simulating the background counts tells us to expect only $<$1 chance coincidence within the 5$^{\prime\prime}$ matching radius.  Therefore, at least 75 of these matches are very likely genuine associations.  While the majority of the newly-detected quasars have sub-mJy peak flux densities, five of them have flux densities above 1 mJy, and therefore should have been seen by the FIRST survey.  In three of the cases, however, the source has a double-lobed morphology, and FIRST's resolution was too poor to resolve the core from one of the radio lobes.  The position of the FIRST source was therefore shifted further than 5$^{\prime\prime}$ from the optical center.  In the other two cases, the FIRST cutout server shows that the sources were clearly not bright enough to appear in the FIRST catalog, implying that the peak flux density of the quasar has brightened in the high resolution data due to variability.  All 76 quasars are therefore newly-discovered radio quasars, and a list of the sources is presented in Table \ref{tab:NEWqso}. 

\begin{figure}
\centering
\includegraphics[scale=0.48]{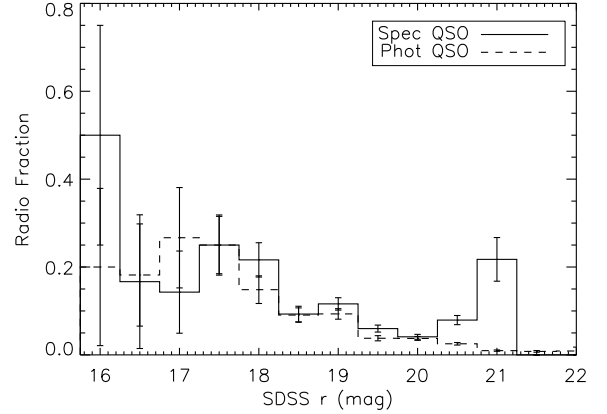}
\caption{Fraction of quasars that are radio sources as a function of SDSS r-band magnitude.  The solid line represents the spectroscopically-confirmed quasars, while the dashed line shows the photometrically-selected quasars.  Error bars were calculated using the formal binomial distribution: $\sigma$($f$) $=$ $[f(1-f)/N]^{0.5}$. }
\label{fig:QSOradiofract}
\end{figure}

As a further comparison, we also matched our radio source catalog to a larger catalog of photometrically-selected quasars based on the SDSS DR6 \citep{Richards:2009p67}.  There are $\sim$13,500 photometrically-selected quasars in the area covered by our data, almost 3.5 times the number of spectroscopically-confirmed quasars.  Despite this fact, the fraction of radio sources that match to quasars only increases to 2.0\% ($\pm$ 0.1\%), a factor of 1.2. Figure \ref{fig:QSOradiofract} shows the fraction of quasars that are radio sources as a function of SDSS r-band magnitude for both the spectroscopic and photometric samples.  The two samples are consistent above r $=$ 20, lending credence to the reliability of the photometrically-selected sample in that magnitude range.  Below r $=$ 20, we see a statistically-significant deficit in the radio fraction of the photometric sample with respect to the spectroscopic sample.  As the peak of the photometrically-selected quasar magnitude distribution lies in this range, this seemingly small difference explains the large discrepancy in the total percentages referenced above.  

We believe that this difference can be explained, at least in part, by the efficiency of the SDSS target selection algorithms.  Targets may be selected for a number of reasons, including stellar objects within 2$^{\prime\prime}$ of a FIRST source, ROSAT sources with quasar colors or that are also radio sources, and, perhaps most notably, in an open category of serendipitously selected targets including objects coincident with FIRST sources but fainter than the equivalent in quasar target selection (and not restricted to point sources).  As a result, the photometric catalog would mainly  fill in those objects without radio detections, which are presumably much more numerous.  This would explain why the total percentage of photometrically-selected quasars that are radio sources (based on our catalog) is only 2.7\% ($\pm$ 0.01\%).  Further investigation into the radio and optical properties of the entire quasar sample will be undertaken in an upcoming paper.


\section{PUBLIC ACCESSIBILITY}
\label{public}
All images are available in searchable format via a link on the FIRST website\footnote{http://third.ucllnl.org/cgi-bin/stripe82cutout}.  The search parameters required are the same as those required by FIRST and are described online. The resultant cutout will be shown alongside the corresponding cutout from the FIRST survey.  The Stripe~82 catalog is also available online in a searchable format\footnote{http://sundog.stsci.edu/cgi-bin/searchstripe82}.  In addition, an option has been added to the FIRST catalog search page to view the corresponding Stripe~82 image in the two non-contiguous areas of coverage.  The full Stripe~82 catalog will be available for download shortly from the same website.    

\acknowledgements
JAH acknowledges the support of NRAO Grant GSSP08-0034, a UC Davis Graduate Block Grant Fellowship, and Grant HST-GO-10412.03-A from the Space Telescope.  RHB acknowledge the support of the National Science Foundation under grant AST 00-98355.  The work by RHB was partly performed under the auspices of the U.S. Department of Energy by Lawrence Livermore National Laboratory under Contract DE-AC52-07NA27344.  RLW acknowledges the support of the Space Telescope Science Institute, which is operated by the Association of Universities for Research in Astronomy under NASA contract NAS5-26555.   GTR acknowledges support from an Alfred P. Sloan Research Fellowship.

The National Radio Astronomy Observatory is a facility of the National Science Foundation under cooperative agreement by Associated Universities, Inc.

\end{document}